\documentclass[twocolumn,trackchanges]{aastex701}

\begin{document}
\title{{Discovery of Changing-Look behavior in AGN NGC 3822: A long-term multiwavelength study}}
\newcommand\nustar{{\it NuSTAR}}
\newcommand\xmm{{\it XMM-Newton}}
\newcommand\nicer{{\it Nicer}}
\newcommand\XMM{{\it XMM}}
\newcommand\suz{{\it Suzaku}}
\newcommand\Swift{{\it Swift/XRT}}
\newcommand\swift{{\it Swift}}
\newcommand\s{{\rm~s}}
\newcommand\ks{{\rm~ks}}
\newcommand\mhz{{\rm~mHz}}
\newcommand\mpc{{\rm~Mpc}}
\newcommand\kpc{{\rm~kpc}}
\newcommand\pc{{\rm~parsec}}
\newcommand\hz{{\rm~Hz}}
\newcommand\kev{{\rm~keV}}
\newcommand\ev{{\rm~eV}}
\newcommand\xiunit{\ifmmode {\rm~erg\s}$^{-1}$ \else ~erg~cm~s$^{-1}$\fi}
\newcommand\kms{\ifmmode {\rm~km\ s}$^{-1}$ \else ~km s$^{-1}$\fi}
\newcommand\Hunit{\ifmmode {\rm~km\ s}$^{-1}$\ {\rm Mpc}$^{-1}$
	\else ~km s$^{-1}$ Mpc$^{-1}$\fi}
\newcommand\cts{\ifmmode {\rm~count\ s}$^{-1}$ \else ~count s$^{-1}$\fi}
\newcommand\ergsec{\ifmmode {\rm~erg\ s}$^{-1}$ \else
	~erg s$^{-1}$\fi}
\newcommand\funit{\ifmmode {\rm~erg\ s}$^{-1}$\;{\rm cm}$^{-2}$ \else
	~erg s$^{-1}$ cm$^{-2}$\fi}
\newcommand\phflux{\ifmmode {\rm~photon\ s}$^{-1}$\;{\rm cm}$^{-2}$
	\else   ~photon s$^{-1}$ cm$^{-2}$\fi}
 \newcommand\normflux{\ifmmode {\rm~photons keV$^{-1}$};{\rm cm}$^{-2}$
	\else   ~photons keV$^{-1}$ cm$^{-2}$ s$^{-1}$\fi}
    \newcommand\flux{\ifmmode {\rm~erg\ s}$^{-1}$\;{\rm cm}$^{-2}$
    \else ~erg cm$^{-2}$ s$^{-1}$\fi}
\newcommand\efluxA{\ifmmode {\rm~erg\ s}$^{-1}$\;{\rm cm}$^{-2}$\;{\rm
		\AA}$^{-1}$ \else ~erg s$^{-1}$ cm$^{-2}$ \AA$^{-1}$\fi}
\newcommand\efluxHz{\ifmmode {\rm~erg\ s}$^{-1}$\;{\rm cm}$^{-2}$\;{\rm
		Hz}$^{-1}$ \else ~erg s$^{-1}$ cm$^{-2}$ Hz$^{-1}$\fi}
\newcommand\cc{\ifmmode {\rm~cm}$^{-3}$ \else cm$^{-3}$\fi}
\newcommand\cs{\ifmmode {\rm~cm}$^{-2}$ \else cm$^{-2}$\fi}
\newcommand\FWHM{\ifmmode {\rm~FWHM} \else ${\rm~FWHM}$\fi}
\newcommand\Msun{\ifmmode M_{\odot} \else $M_{\odot}$\fi}
\newcommand\Lsun{\ifmmode L_{\odot} \else $L_{\odot}$\fi}
\newcommand\ltsim{\raisebox{-.5ex}{$\;\stackrel{<}{\sim}\;$}}
\newcommand\gtsim{\raisebox{-.5ex}{$\;\stackrel{>}{\sim}\;$}}
\newcommand\hbeta{\ifmmode {\rm H}\beta \else H$\beta$\fi}
\newcommand\halpha{\ifmmode {\rm H}\alpha \else H$\alpha$\fi}
\newcommand\Kalpha{\ifmmode {\rm K}\alpha \else K$\alpha$\fi}
\newcommand\nh{\ifmmode N_{\rm H} \else N$_{\rm H}$\fi}
\newcommand{\OIII}{[\mathrm{O\,III}]~\lambda\lambda4959, 5007}
\newcommand{\OI}{[\mathrm{O\,I}]~\lambda6300}
\newcommand{\NII}{[\mathrm{N\,II}]~\lambda\lambda6548,6584}
\newcommand{\SII}{[\mathrm{S\,II}]~\lambda6716}
\author[orcid=0009-0009-8761-1798]{Narendranath Layek}
\affiliation{Astronomy and Astrophysics Division, Physical Research Laboratory, Navrangpura, Ahmedabad - 380009, Gujarat, India}
\affiliation{Indian Institute of Technology Gandhinagar, Palaj, Gandhinagar - 382055, Gujarat, India}
\email[show]{narendral@prl.res.in, narendranathlayek2017@gmail.com (NL)}
\author[0000-0003-3840-0571]{Prantik Nandi}
\affiliation{Indian Centre for Space Physics, Netaji Nagar, Kolkata, 700099, India}
\email{prantiknandiphy@gmail.com}
\author[0000-0003-2865-4666]{Sachindra Naik}
\affiliation{Astronomy and Astrophysics Division, Physical Research Laboratory, Navrangpura, Ahmedabad - 380009, Gujarat, India}
\email{snaik@prl.res.in}
\author[0000-0002-9680-7233]{Birendra Chhotaray}
\affiliation{Astronomy and Astrophysics Division, Physical Research Laboratory, Navrangpura, Ahmedabad - 380009, Gujarat, India}
\email{rsbirendra786@gmail.com}
\author[0000-0001-7500-5752]{Arghajit Jana}
\affiliation{Instituto de Estudios Astrof\'isicos, Facultad de Ingenier\'ia y Ciencias, Universidad Diego Portales, Av. Ej\'ercito Libertador 441, Santiago, Chile}
\affiliation{Department of Physics, SRM University-AP, Amaravati, 522240, Andhra Pradesh, India}
\email{argha0004@gmail.com}
\author[0009-0002-7307-3561]{Priyadarshee P. Dash}
\affiliation{Astronomy and Astrophysics Division, Physical Research Laboratory, Navrangpura, Ahmedabad - 380009, Gujarat, India}
\affiliation{Indian Institute of Technology Gandhinagar, Palaj, Gandhinagar - 382055, Gujarat, India}
\email{ppdash@prl.res.in}
\author[0000-0003-0071-8947]{Neeraj Kumari}
\affiliation{Indian Institute of Astrophysics, Block II, Koramangala, Bangalore, 560034, India}
\affiliation{INAF-IASF Palermo, Via Ugo La Malfa 153, I-90146 Palermo, Italy}
\email{neerjakumari108@gmail.com}
\author[0000-0002-4998-1861]{C. S. Stalin}
\affiliation{Indian Institute of Astrophysics, Block II, Koramangala, Bangalore, 560034, India}
\email{stalin@iiap.res.in}
\author[0009-0004-1054-2812]{Srikanth Bandari}
\affiliation{Indian Institute of Astrophysics, Block II, Koramangala, Bangalore, 560034, India}
\email{bandari.srikanth@iiap.res.in}
\author[0000-0002-4024-956X]{S. Muneer}
\affiliation{Indian Institute of Astrophysics, Block II, Koramangala, Bangalore, 560034, India}
\email{muneers@iiap.res.in}

\begin{abstract}
We present a comprehensive long-term multi-wavelength study of the active galactic nucleus (AGN) NGC~3822, based on 17 years (2008–2025)  of X-ray, ultraviolet (UV), and optical observations. The dataset includes observations from {\it Swift, XMM-Newton}, and {\it NuSTAR}, the Very Large Telescope, and the Himalayan Chandra Telescope. Our multiwavelength light curve analysis reveals flux variations across X-ray to optical/UV bands, with an increased variability amplitude at shorter wavelengths. X-ray spectral analysis indicates the presence of intrinsic absorption during the 2016 and 2022 observations; however, this absorption disappeared before and after these epochs. The presence and absence of the absorber are attributed to clouds moving in and out of the line of sight. During the long-term monitoring period, the bolometric luminosity of the source varies between ($1.32-17)\times10^{43}$ erg s$^{-1}$. Optical spectroscopic monitoring reveals changing-look (CL) behaviour in NGC~3822, characterized by the appearance and disappearance of broad emission lines (BELs). These CL transitions are associated with changes in the Eddington ratio rather than changes in the obscuration. The BELs appear only when the Eddington ratio is relatively high ($\sim 3.8\times10^{-3}$) and disappear when it drops to a lower value ($\sim 0.9\times10^{-3}$).
\end{abstract}

\section {Introduction}
Active galactic nuclei (AGNs) are among the most luminous and energetic sources in the universe. This extreme luminosity of the AGNs is understood to arise from the accretion of matter onto supermassive black holes (SMBHs) residing at the centre of host galaxies~\citep{Rees1984}. The AGNs emit radiation across the entire electromagnetic spectrum, ranging from radio waves to high-energy $\gamma$-rays. The optical/UV emission is believed to originate from an optically thick accretion disk~\citep{Shakura1973}. Viscous dissipation on the disk generates heat, which is then radiated in the optical/UV regime~\citep{Sun1989}. X-rays in radio-quiet AGNs originate primarily from a cloud of hot electrons, termed as corona, located within a distance of 3--10$r_g$ above the central black hole~\citep{Fabian2009,Cackett2014,Kara2014,Fabian2015}. A fraction of optical/UV photons from the accretion disk are inverse Compton scattered by the electrons in the hot, optically thin corona and upscattered into X-rays~\citep{Sunyaev&Titarchuk1980, Sunyaev1985, 1991ApJ...380L..51H,Haardt1993H,1994ApJ...428L..13N, 1995ApJ...455..623C}. The X-ray spectrum of AGNs is typically characterized by a power-law continuum, which often exhibits a high-energy exponential cutoff, usually around a few hundred kilo-electronvolts~\citep{Sunyaev&Titarchuk1980}. This feature is directly related to the temperature and optical depth of the plasma of hot electrons responsible for the power-law continuum. However, the properties of the corona, such as its size, geometry, and position, are still subjects of debate. In the optical/UV regime, AGNs are generally classified as type~1 or type~2 based on the widths of the optical emission lines. Type~1 AGNs show both broad emission lines (BELs; full widths at half maximum~$>1000$\kms) originating in the broad line region (BLR) and narrow emission lines (NELs; full widths at half maximum~$<1000$\kms) originating in the narrow line region (NLR). However, the type~2 AGNs show only NELs in their UV/optical spectra. Depending on the width of the emission lines and the relative strength of the BELs to the NELs, finer classifications (type~1.5, 1.8, and 1.9) are used~\citep{Osterbrock1981,Winkler1992}. Typically, in type~1, the broad line dominates over the narrow line, while in type~1.5, the narrow lines are more noticeable. In Seyfert~1.8, only broad \halpha~ and faint broad \hbeta~ lines are present (\citealt{Cohen1986}), and in Seyfert~1.9, only the broad \halpha~ line is visible. In X-rays, on the other hand, AGNs are classified on the basis of their obscuration properties, and in particular by their line-of-sight hydrogen column density ($N_{\rm H}$). If $N_{\rm H}>10^{22}$~\cs, the AGNs are called obscured AGNs, while AGNs with $N_{\rm H}<10^{22}$~\cs~ are classified as unobscured AGNs. Generally, type~1 AGNs are found to be unobscured, and the type~2 AGNs are obscured \citep{Ricci2017,Koss2017,Oh2022}. The classification of AGNs can be described using the simplified unification model~\citep{Antonucci1993} based on the line of sight of the obscuring torus. In type~2 AGNs, the dusty torus surrounding the SMBH at a distance of parsecs to tens of parsecs obscures the BLR, while in type~1 AGNs, the observer has a direct view of the BLR.

In recent years, several tens of subclasses of AGNs have been discovered, which show dramatic optical and X-ray spectral variabilities on timescales ranging from months to decades~\citep{Parker2016,Yang2018,jana2025}. These are known as changing-look AGNs (CL-AGNs) and are currently an open issue in AGN physics. In the recent review article by \cite{Ricci2023}, the CL-AGNs are classified into two classes based on their properties in the UV/optical and X-ray regimes. In optical/UV, these objects show the appearance or disappearance of the broad optical emission lines, switching from type~1 (or type~1.2/1.5) to type~2 (or type~1.8/1.9) and vice versa on a time scale of months to decades~\citep{Ricci2023}, and are usually considered as "changing-state" AGNs (CSAGNs). In X-rays, a different type of changing-look event is observed with rapid variability on $N_{\rm H}$ on a timescale of hours to years, known as "changing-obscuration" AGNs (COAGNs). Over the years, many AGNs, such as NGC~4151~\citep{Puccetti2007}, Mrk~590~\citep{Shappee2014}, NGC~3516~\citep{Ili2020}, Mrk~1018~\citep{Cohen1986}, NGC~1566~\citep{Oknyansky2019,Jana2021}, and many more, have been found to show CL transitions on a timescale of months to decades. The origin of the changing-state (CS) and changing-obscuration (CO) events remains unclear, and several models have been proposed to explain them. In general, COAGNs are associated with obscuration caused by clumpiness of the BLR or the presence of a circumnuclear torus of molecular gas and dust (\citealt{Nenkova2008,Nenkovab2008,Ricci2016,Jana2020,Jana2022}). CSAGNs are believed to be caused by changes in accretion rate, which can be attributed to local disk instabilities \citep{Stern2018,Noda2018} or major disk perturbations, such as tidal disruption events (TDEs; \citealt{Merloni2015,Ricci2020}).

The AGN NGC~3822 is a nearby AGN (z = 0.019) with a central supermassive black hole mass of $M_{\rm BH}=2.70\times10^7~M_\odot$ \citep{Chen2024}. From the optical observations, NGC~3822 was initially classified as Seyfert~2~\citep{Moran1994} in 1994. Later, it was found that this source exhibited broad $H_{\alpha}$ variability within one year (from 1994 to 1995) and showed Seyfert~1.9 characteristics in 1995~\citep{Moran1996}. In 2022, a spectrum was obtained using the Supernova Integral Field Spectrograph (SNIFS) as part of the Spectral Classification of Astronomical Transients (SCAT) program, which revealed a weak blue continuum along with broad ($\sim$5000 km/s) Balmer and He~I lines at the host galaxy redshift. Based on these broad emission lines, \citet{Hinkle2022} classified NGC~3822 as a Seyfert~1 AGN. They reported this AGN as a CL classification. Although there has been some past optical spectroscopic investigation, the source has not been thoroughly studied in the X-ray domain. From a sample study of IRAS sources detected in the ROSAT All-Sky Survey,~\cite{Moran1994} estimated that the X-ray luminosity of this source is approximately $\sim1.86\times10^{42}$\ergsec~ in the 0.1-2.4 keV range.

In this work, we present our comprehensive findings of the long-term ($\sim$16 yr; from 2008 to 2024) multi-wavelength observations of NGC~3822 from various observatories. The paper is structured in the following way. Section~\ref{sec:obs_data} provides an overview of the observational data and describes the procedures used for data reduction. The results are presented in detail in Section~\ref{sec:results}. Then, we discuss our key findings in Section~\ref{sec:discussion}, and finally, our conclusions are summarized in Section~\ref{sec:conclusions}.
\begin{table*}

	\centering
	\caption{Details of the multiwavelength observations of NGC~3822 used in this work.}
	\label{tab:obs}
	\begin{tabular}{l c c c c c } 
	\hline
    \hline
    Observation Band & Telescope    &   Obs. Date          & Obs. ID      &     Exposure & Short ID\\
                     &              & (yyyy-mm-dd)         &              &      (s)    &       \\
    \hline
    X-ray and UV/Optical    & \swift-XRT/UVOT &  2008-07-23    &  00036986001     & 2000             & XRT1   \\
                    &  \xmm-EPIC-pn/OM & 2010-06-01    &  0655380101      & 14000           & XMM   \\ 
                    & \swift-XRT/UVOT & 2013-07-26     & 00036986002      & 1000             & XRT2 \\
                    & --              & 2015-03-24     & 00085582004      & 5170             & XRT3 \\
                     & --              & --2015-07-17     & --00085582007      & --          &  \\
                    & --              & 2016-01-12     & 00080667001      & 2110             & XRT4 \\
                    & --              & --2016-01-15     & --00080667002      & --           & \\
                    & \nustar-FPMA/FPMB & 2016-01-12   & 60061332002      & 21230            & NU1 \\
                    & \swift-XRT/UVOT   & 2022-03-30   & 00037008004      & 5640             & XRT5a \\
                    & --                & --2022-05-28   & --00037008008      & --               &  \\
                    & \nustar-FPMA/FPMB & 2022-06-04   & 90801611002      & 21300            & NU2 \\
                    
                    & \swift-XRT/UVOT   & 2022-06-12   & 00037008009      & 3300             & XRT5b \\
                    & --                & --2022-06-26   & 00037008011      & --             &  \\
                    & --                & 2022-07-07   & 00037008013      & 4520             & XRT5c \\
                    & --                & --2022-07-17   & 00037008014      & --               &  \\
                    & --                & 2022-11-02   & 00037008015      & 10650            & XRT5d \\
                    & --                & --2022-11-29   & 00037008021      & --               &  \\
                    & --                & 2023-06-12   & 00037008022      & 7200             & XRT6 \\
                    & --                & --2023-12-22   & 00036986017      & --               &  \\
                    & --                & 2024-04-14   & 00037008024      & 11510              & XRT7 \\
                    & --                & --2024-07-13   & 00037008032      & --               &  \\
\hline    
Optical             & VLT/X-Shooter      &2018-06-11                &ADP.2018-06-11T11                  & 480                 &XS1   \\
                    & --                &2022-06-17                &ADP.2022-06-17T14                  & 480                 &XS2    \\
                    & --                &2022-07-11                &ADP.2022-07-11T14                  & 480                 &XS3     \\
                    & HCT/HFOSC         &2024-03-06                &HCT-2024-C1-P43                    &700               &HCT1  \\
                    & --                &2024-04-01                &--                                 &2100              &HCT2   \\
                    & --                &2024-05-02                &HCT-2024-C2-P7                     &2100              &HCT3    \\
                    & --                &2024-06-02                &--                                 &2100              &HCT4  \\
                    & --                &2025-01-01                &HCT-2025-C1-P11                    &2100              &HCT5   \\
                    & --                &2025-03-03                &HCT-2025-C1-P38                    &1400              &HCT6    \\
\hline

	\end{tabular}
 \end{table*}
\section{Observations and Data Reduction}
\label{sec:obs_data}
\subsection{X-ray, UV and Optical Continuum Observations}
In this work, we used publicly available archival data from \swift,~\xmm, and \nustar~ observatories during the period between 2008 and 2024. All data sets are reduced and analyzed using the {\tt HEAsoft v6.30.1} package. The details of each observatory and the standard data reduction procedure are described in the following subsections. 
\subsubsection{Swift}
\label{sec:swift_red}
\swift~ \citep{2004ApJ...611.1005G} is a multiwavelength observatory operating in the optical/UV to X-ray wavebands. NGC~3822 was observed by \swift~ over several epochs from 2008 to 2024. We reduced the data of \swift/XRT observations using the online tool “Swift Build XRT products routine”\footnote{\url{http://swift.ac.uk/user objects/}}. The tool processes and calibrates the data and produces the final spectra and light curves of NGC~3822 in two modes, e.g., window timing (WT) and photon counting (PC) modes. However, due to the low signal-to-noise ratio (SNR), we combined these observations into seven distinct groups labeled XRT1, XRT2, XRT3, XRT4, XRT5, XRT6, and XRT7. However, for the 2022 observation (XRT5), a subsequent categorization (XRT5a, XRT5b, XRT5c, and XRT5d) is made due to the relatively higher SNR during this observation period (see Table~\ref{tab:obs}). Due to low SNR, we fitted all the XRT spectra using Cash statistics after binning the spectra to have at least 5 counts per bin to avoid possible issues related to empty bins in $\tt XSPEC$.

There are six filters in the optical/UV band of ~\swift/UVOT, which are V (5468~\AA), B (4392~\AA), U (3465~\AA), UVW1 (2600~\AA), UVM2 (2246~\AA), and UVW2 (1928~\AA) bands. The UVOT data are reduced from the level~II image by using the tool {\tt UVOTSOURCE}. To obtain the source counts, we assume a circular region of a 5 arcsec radius centered on the source position, whereas a circular region with a 20 arcsec radius away from the source position is considered for background counts. The observed optical/UV fluxes are corrected for reddening and Galactic extinction using the reddening coefficient $E(B-V)=0.0479$ obtained from the Infrared Science Archive \footnote{\url{http://irsa.ipac.caltech.edu/applications/DUST/}} and $R_{V}=A_{V}/E(B-V)=3.1$ following ~\cite{Schlafly2011}. The Galactic extinction ($\rm A_{\lambda}$) values calculated for V, B, U, UVW1, UVM2, and UVW2 bands are 0.15, 0.19, 0.24, 0.34, 0.45, and 0.39, respectively, assuming the extinction model of ~\cite{Fitzpatrick2007}. The corrected UVOT fluxes are listed in Table~\ref{tab:UV/optical_flux}.

\subsubsection{XMM-Newton}
\label{sec:xmm_red}
NGC~3822 was observed with \xmm~\citep {2001A&A...365L...1J} only once in June 2010. Data from the {\it European Photon Imaging Camera } (EPIC;~\citealt{struder2001}) and {\it Optical Monitor }(OM;~\citealt{Mason2001}) instruments are used in the present work. To process the raw data from EPIC-pn, we use Science Analysis System ({\tt SAS v18.0.0}). Only unflagged events with {\tt PATTERN $\leq$ 4} are considered in our analysis. We excluded flaring events from the data by choosing appropriate {\tt GTI} files. The observed data are corrected for the photon pile-up effect by considering an annular region with outer and inner radii of 30 and 5 arcsecs, respectively, centred at the source coordinates while extracting the source events. We use a circular region of 60 arcsec radius, away from the source position, for the background products. The response matrices ({\it arf} and {\it rmf}) are created using the SAS tasks {\tt ARFGEN} and {\tt RMFGEN}. The \xmm~spectra are binned using the GRPPHA task to ensure a minimum of 25 counts per energy bin.

NGC~3822 was observed with the Optical Monitor (OM) in imaging mode with only UVM2 (2310~\AA) filter. OM data are processed using the task {\tt OMCHAIN}. We obtain the background corrected count rate from the source list for the OM filter and convert the count rate into the respective flux density~\footnote{\url{https://www.cosmos.esa.int/web/xmm-newton/sas-watchout-uvflux}}. The flux density is then corrected for Galactic extinction and reported in Table~\ref{tab:UV/optical_flux}. 

\subsubsection{NuSTAR}
\label{sec:Nustar_red}
NuSTAR~\citep{2013ApJ...770..103H} is a hard X-ray focusing telescope with two identical focal plane modules (FPMA and FPMB), operating in the 3-79 keV energy range. NGC~3822 was observed by NuSTAR simultaneously with \swift~ in January 2016 and June 2022. We use the standard NuSTAR Data Analysis Software {\tt (NuSTARDAS v2.1.2\footnote{\url{https://heasarc.gsfc.nasa.gov/docs/nustar/analysis/}})} package to extract data. The standard {\tt NUPIPELINE} task with the latest calibration files CALDB \footnote{\url{http://heasarc.gsfc.nasa.gov/FTP/caldb/data/nustar/fpm/}} is used to generate cleaned event files. The {\tt NUPRODUCTS} task is utilized to extract the source and background spectra and light curves. We consider circular regions of radii 60 and 120 arcsecs for the source and background products, respectively. The circular region for the source is selected with the centre at the source coordinates, and the region for the background is chosen far away from the source to avoid contamination. Due to the low SNR ratio, the \nustar~ spectra are grouped into 10 counts per energy bin.
\subsection{Optical Spectroscopy}
The optical data utilized in this study were acquired through approved observation (PI: N. Layek) from the Himalayan Chandra Telescope (HCT), as well as from publicly available archival data sets provided by ESO~\footnote{\url{https://archive.eso.org/scienceportal/home}}. The following section provides details on each telescope and the standard data reduction procedures. 
\subsubsection{HCT}
We observed NGC~3822 from 03 July 2024 to 03 July 2025 using the Hanle Faint Object Spectrograph and Camera (HFOSC) instrument mounted on the 2~m Himalayan Chandra Telescope (HCT) of the Indian Astronomical Observatory (IAO), located at Hanle, Ladakh, India~\citep{Cowsik2002}. Spectra were taken with Grisms~7 (Gr7) and 8 (Gr8) with different slit options. Gr7 has a spectral resolution of 1330 with a wavelength coverage of 3800--6840~\AA. For Gr8, the wavelength coverage is 5800--8350~\AA, and the spectral resolution is 2190. We acquired flat, bias, and calibration lamp frames, along with standard and target star exposures. Data reduction was performed using self-developed data analysis routines in the {\it Python} and NOAO Image Reduction and Analysis Facility (IRAF; ~\citealt{Tody1986}), following the same procedure as described in \citet{2024MNRAS.534.2830C}. In the first step, bias subtraction, cosmic-ray removal ~\citep{vanDokkum2001}, and flat-fielding are done for all frames. Halogen lamp frames are used for flat-fielding the images. Then, one-dimensional spectra are extracted after removing the mean sky background extracted from both sides of the source spectrum. In the second step, wavelength calibration was performed using FeAr and FeNe calibration lamps for Gr7 and Gr8, respectively. The instrument response function was generated with Feige56 as the standard star for Gr7 and Gr8. Subsequently, science-ready spectra were produced by applying this response function to the target AGN (NGC~3822) spectra.
\subsubsection{VLT}
NGC~3822 was observed with the X-Shooter spectrograph mounted on the Very Large Telescope (VLT)~\citep{Vernet2011}. The X-Shooter is an intermediate resolution slit spectrograph with a resolution ranging from R$\sim$4000--17,000, and covers a wavelength range from 3000 to 25000 \AA, divided into three arms: UV-Blue (UVB), visible (VIS), and near-infrared (NIR). In this work, we used data from the UVB and VIS arms at three epochs: July 2018, July 2022, and May 2022. Observations were carried out with slit dimensions and spectral resolutions of 1.6~arcsec~$\times$~11~arcsecs in the UVB arm (R = 3200) and 1.5~arcsec~$\times$~11~arcsecs in the VIS arm (R = 5000). We accessed the processed science data provided by the ESO observatory\footnote{\url{https://archive.eso.org/scienceportal/home}}.
\begin{figure*}
\centering
\hspace*{1cm}
\includegraphics[trim={10 9.5cm 2cm 0},scale=0.80]{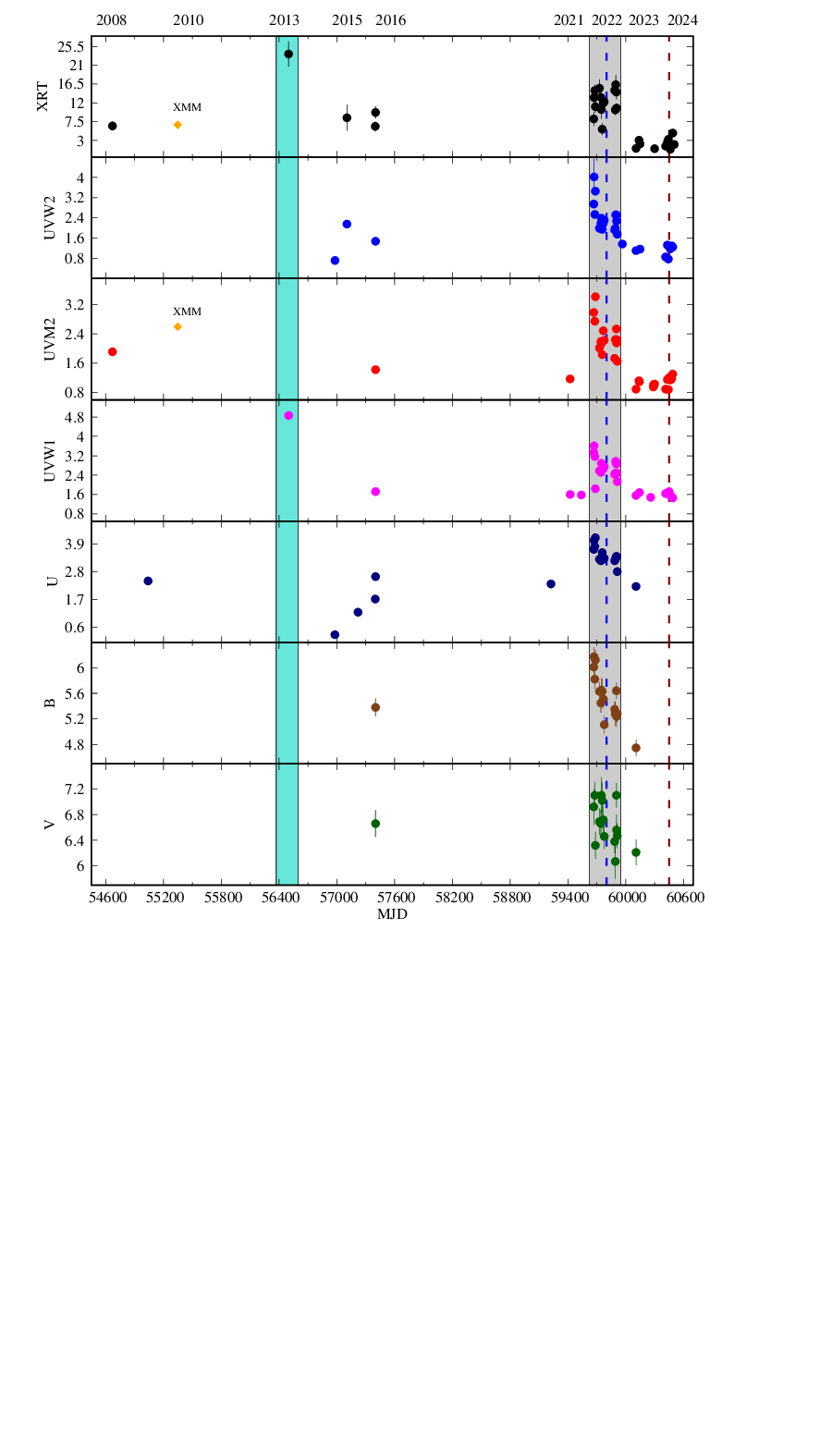}
\includegraphics[trim={0 9.5cm 8cm 0},scale=0.80]{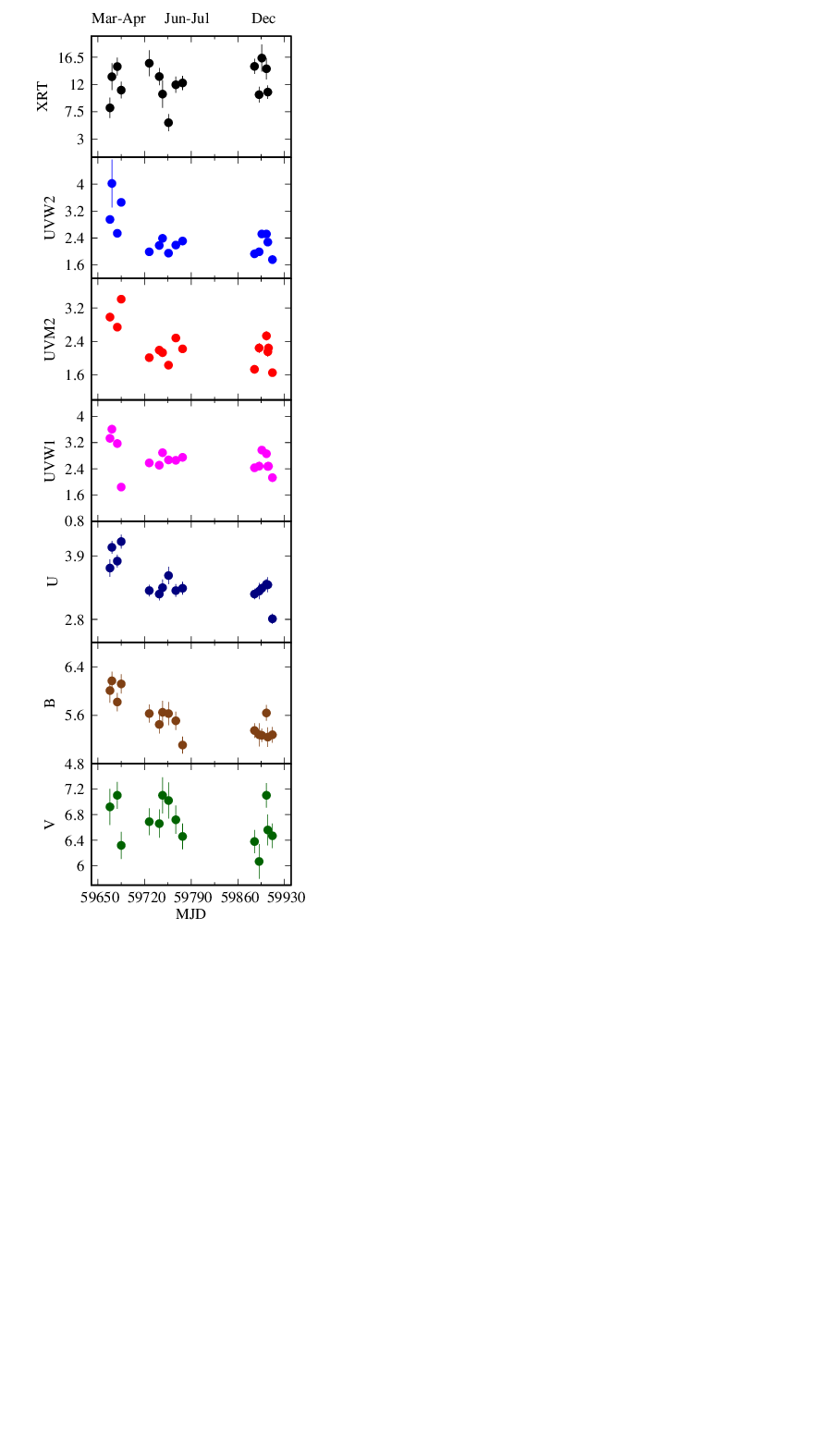}  
\caption{ {\it Left:} Temporal variation of X-ray (0.3--10 keV range), UV, and optical continuum flux obtained from the ~\swift~(XRT \& UVOT) and \xmm~(EPIC \& OM) observations for the years 2008 to 2024. The shaded areas in cyan and grey indicate the observations for the years 2013 and 2022, respectively, during which the source exhibited significantly enhanced flux levels across all bands compared to other epochs. The dotted blue vertical line indicates the epoch when BELs are detected, while the red line marks the epoch when BELs are not detected. {\it Right:} Zoomed-in view of the 2022 monitoring campaign, covering the period from March 2022 to December 2022. The X-ray flux is given in the unit of $10^{-12}$~\flux~and the optical/UV monochromatic flux is in the unit of $10^{-15}$\efluxA. }
\label{fig:all_flux}
\end{figure*}
\section{Results}
\label{sec:results}
\subsection{X-ray, UV, and optical continuum flux variations}
\label{sec:Fvar}
We present the long-term (2008--2024) multiwavelength light curves of NGC~3822 in Figure~\ref{fig:all_flux} (left panel). In this figure, we used UV and optical continuum fluxes obtained from ~\swift-UVOT and ~\xmm-OM observations, along with 0.3-10 keV X-ray fluxes from ~\swift/XRT and ~\xmm/EPIC-pn observations. The 0.3-10 keV X-ray fluxes were retrieved by fitting the X-ray spectra of NGC~3822 with a power-law model. All flux measurements are listed in Table~\ref{tab:UV/optical_flux} and Table~\ref{tab:X_ray_flux}. These multiwavelength observations were used to study the long-term flux variations across different wavelength bands in NGC~3822. From the multi-wavelength light curve, it is evident that the source found to be in an X-ray high state during the 2013 observation, with a total X-ray (0.3-10 keV) flux of $23.71\pm3.03\times10^{-12}$\flux. Before and after this period, in 2010 and 2015, the source transitioned into a low X-ray state, with 0.3-10 keV X-ray fluxes of $6.75\pm0.18\times10^{-12}$~\flux and $8.44\pm3.13\times10^{-12}$\flux, respectively (Table~\ref{tab:X_ray_flux}). By 2022, the source had returned to a high X-ray state, with an average total X-ray flux of $12 \times10^{-12}$\flux and corresponding minimum and maximum fluxes of $5.70 \times10^{-12}$~\flux ~and $16.37 \times10^{-12}$~\flux, respectively. Along with X-ray, a significant increase in flux was also observed in the optical/UV band, as shown in Figure~\ref{fig:all_flux}. During the 2013 observation, in the UVW1 band, the flux increased to approximately $4.8\times10^{-15}$\efluxA, nearly three times the flux measured in 2016. After that, in 2022, a significant enhancement of flux was observed in the optical/UV regime, followed by a gradual decline during the 2023-2024 observations. The right panel of Figure~\ref{fig:all_flux} displays the X-ray, UV, and optical \swift~light curves for the detailed campaign conducted in 2022. 

In Figure~\ref{fig:all_flux}, it is observed that the source exhibits variability across all observed bands, from optical/UV to X-rays. This result allows us to investigate the physical mechanisms of the inner accretion flow in NGC~3822. To quantify variability across different wavebands, we calculated the fractional variability amplitude ($F_{var}$) for the long-term monitoring data spanning 2008 to 2024 and the short-term monitoring campaign in 2022. The fractional variability is given by the relation
\begin{equation}
F_{\rm var}=\sqrt{\frac{{\sigma^2_{\rm XS}}}{\mu^2}} 
\end{equation}
where $\sigma^2_{\rm XS}$ is the excess variance (\citealt{1997ApJ...476...70N}; \citealt{2002ApJ...568..610E}), used to estimate the intrinsic source variance and given by
\begin{equation}
\sigma^2_{\rm XS}=\sigma^2- \frac{1}{N} \sum_{i=1}^N \sigma^2_{i}
\end{equation}

where $\sigma$ is the variance of the light curve, $\sigma_{i}$ is the error in each flux measurement, $\mu$ is the mean flux measurement, and N is the number of flux measurement. The normalized excess variance is defined as $\sigma^2_{\rm NXS}=\sigma^2_{\rm XS}/\mu^2$. The uncertainties in $\sigma^2_{\rm NXS}$ and $F_{\rm var}$ are estimated as described in \cite{2003MNRAS.345.1271V} and \cite{2012ApJ...751...52E}. We calculate maximum ($F_{max}$), minimum ($F_{min}$) and mean fluxes ($\mu$), peak-to-peak amplitudes $R=F_{max}/F_{min}$, over the observation period from 2008 to 2024. 
For the long-term study, the variability amplitude $\rm F_{var}$ in the X-ray band was found to be $\sim$67$\%$, which gradually decreases in the UV bands from about $\sim$40$\%$ in W2, $\sim$39$\%$ in M2 and $\sim$33$\%$ in W1, followed by a further drop to $\sim$28$\%$ in the optical U band. However, a sudden decline in variability is observed in the V and B bands, where the fractional variability drops significantly to around$\sim$5-3$\%$. During the short-term observations, a similar gradual decrease in the amplitude of the variability was also observed. All variability measurements from long-term and short-term monitoring across different wavebands are summarized in Table~\ref{tab:F_var}.

\begin{table*}
 \setlength{\tabcolsep}{2.0pt}
 \renewcommand{\arraystretch}{1.3} 
    \centering
	\caption{Long-term (2008-2024) and short-term (2022) variability statistics in different bands.} 
	\label{tab:F_var}
     \hspace*{-3.5cm}
	\begin{tabular}{|c| c c c c c  c | c c c c c  c|} 
	  \hline
     \hline
    & \multicolumn{6}{c|}{2008--2024 }& \multicolumn{6}{c|}{2022} \\
\hline
Band     &   $\rm F_{max}$ & $\rm F_{min}$ & $\mu$ &$\rm R $ &$\rm \sigma^{2}_{NXS}$ & $\rm F_{var}$&$\rm F_{max}$ & $\rm F_{min}$ & $\mu$ &$\rm R $ &$\rm \sigma^{2}_{NXS}$ & $\rm F_{var}$\\
      &                 &               &          &      &$(10^{-2})$         &$\%$&                 &               &          &      &$(10^{-2})$         &$\%$\\
     \hline
XRT & 23.71 & 0.9 & 8.22 & 26.34 &$44.61\pm4.18$&$66.80\pm8.70$ &16.37 & 5.70 &  12.24 &2.87 &$4.10\pm1.47$&$20.24\pm5.20$\\
W2  & 4.02 & 0.72 & 1.89 & 5.58 &$16.10\pm1.16$&$40.12\pm5.46$  & 4.02  & 1.76 & 2.43 & 2.28 &$5.02\pm0.91$&$22.41\pm4.45$\\
M2  & 3.41 & 0.88 & 1.70 & 3.87 &$15.84\pm5.90$&$39.80\pm5.10$  & 3.41  & 1.65 & 2.30 & 2.06 &$3.76\pm0.37$&$19.40\pm3.67$\\
W1  & 4.86 & 1.46 & 2.32 & 3.33 &$11.04\pm0.42$&$33.24\pm4.33$  & 3.61 & 1.84 &  2.69 & 1.96 &$2.25\pm0.24$&$15.00\pm2.27$\\
U   & 4.15 & 0.31 & 3.01 & 13.38 &$8.35\pm0.42$&$28.90\pm4.41$  & 4.15 & 2.81 &  3.45 & 1.48 &$0.75\pm0.15$&$8.67\pm1.76$\\
B   & 6.17 & 4.75 & 5.51 & 1.29 &$0.33\pm0.08$&$5.75\pm1.18$    & 6.17 & 5.11 &  5.57 & 1.21 &$0.23\pm0.07$&$4.89\pm1.11$\\
V   & 7.10 & 6.07 & 6.65 & 1.17 &$0.11\pm0.07$&$3.33\pm1.20$    & 7.10 & 6.07 &  6.65 & 1.17 &$0.11\pm0.07$&$3.33\pm1.20$ \\
\hline
     \end{tabular}
\begin{minipage}{15cm}
\small \textbf{Notes:} The X-ray flux is given in the unit of $10^{-12}$~\flux~and the optical/UV monochromatic flux is in the unit of $10^{-15}$\efluxA.
\end{minipage}
\end{table*}
\begin{figure*} 
	\centering
    \hspace*{-0.5cm}
	\includegraphics[trim={0.0cm 4cm 0.5cm 0},scale=1.2]{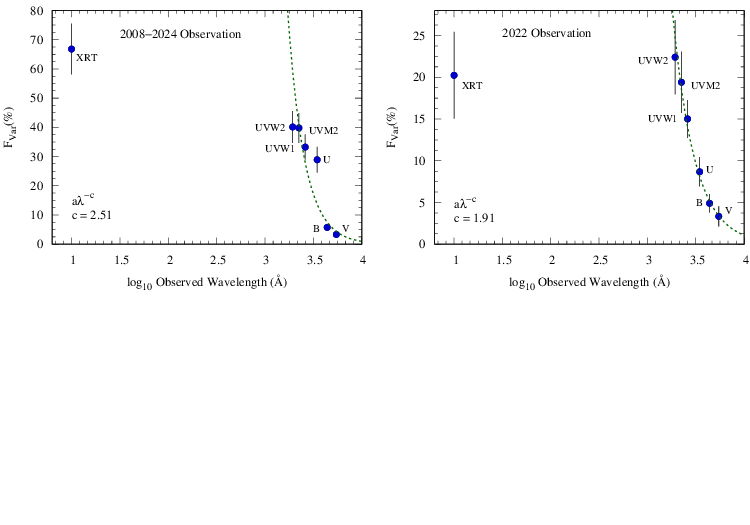} 
    \caption{Fractional variation of the X-ray, UV, and optical continuum bands as a function of wavelength.}
	\label{fig:Fvar} 
\end{figure*}
\begin{table*}
 \setlength{\tabcolsep}{1.8pt} 
 \renewcommand{\arraystretch}{1.2} 
    \centering
	\caption{Best-fit X-ray spectral parameters with model: ${\tt Constant\times Tbabs\times Pcfabs\times Powerlaw}$.} 
	\label{tab:powerlaw_fit}
        \hspace*{-3.5cm}
	\begin{tabular}{l  c c c c c   c c c r} 
	\hline
     \hline
     ID  & MJD    &$N_{H}$&$C_{f}$     &$\Gamma$& $\rm Norm^{PL^\dagger}$&$\rm Satistic/dof$&$\rm logL_{X}$&$\rm log\lambda_{Edd}$&$\rm logL_{Bol}$\\
         &        &($10^{22}cm^{-2}$) &  &        & ($10^{-3}$)            &                  &(\ergsec)       &                   &(\ergsec)                              \\
    \hline
     XRT1& 54670&--&--&$1.55^{+0.17}_{-0.18}$&$0.81^{+0.01}_{-0.01}$&23.83/22&$42.53^{+0.10}_{-0.10}$&$-2.66^{+0.10}_{-0.10}$&$43.63^{+0.10}_{-0.10}$\\
     XMM &55348 &--&--&$1.54^{+0.02}_{-0.02}$&$0.84^{+0.01}_{-0.01}$&405/375&$42.57^{+0.02}_{-0.02}$&$-2.62^{+0.02}_{-0.02}$&$43.67^{+0.02}_{-0.02}$\\
     XRT2& 56499&--&--&$1.53^{+0.15}_{-0.15}$&$2.92^{+0.33}_{-0.35}$&55.19/49&$43.10^{+0.09}_{-0.08}$&$-2.06^{+0.09}_{-0.08}$&$44.23^{+0.09}_{-0.08}$\\
     XRT3&57162&--&--&$1.41^{+0.18}_{-0.18}$&$0.42^{+0.06}_{-0.05}$&28.53/33&$42.35^{+0.10}_{-0.11}$&$-2.84^{+0.10}_{-0.11}$&$43.45^{+0.10}_{-0.11}$\\
     XRT4+NU1& 57339&$1.20^{+1.01}_{-0.71}$&$0.62^{+0.09}_{-0.12}$&$1.80^{+0.11}_{-0.10}$&$1.13^{+0.03}_{-0.02}$&136.78/137&$42.55^{+0.10}_{-0.09}$&$-2.64^{+0.08}_{-0.08}$  &$43.65^{+0.08}_{-0.08}$\\
     
     XRT5a &59697&$1.92^{+1.03}_{-0.68}$&$0.67^{+0.13}_{-0.23}$&$1.93^{+0.28}_{-0.29}$&$2.28^{+0.11}_{-0.82}$&108.95/116&$42.72^{+0.05}_{-0.05}$&$-2.47^{+0.05}_{-0.05}$&$43.82^{+0.05}_{-0.05}$\\
     XRT5b+NU2  &59745&$1.44^{+0.66}_{-0.53}$&$0.73^{+0.09}_{-0.15}$&$1.63^{+0.17}_{-0.17}$&$1.48^{+0.57}_{-0.42}$&148.80/135&$42.73^{+0.03}_{-0.03}$&$-2.46^{+0.03}_{-0.03}$&$43.83^{+0.03}_{-0.03}$\\
     XRT5c &59772&$1.75^{+0.93}_{-0.70}$&$0.77^{+0.11}_{-0.25}$&$1.74^{+0.38}_{-0.41}$&$1.51^{+1.12}_{-0.70}$&80.73/69&$42.67^{+0.07}_{-0.07}$&$-2.52^{+0.07}_{-0.07}$&$43.77^{+0.07}_{-0.07}$\\
     XRT5d&59898&--&--&$1.53^{+0.06}_{-0.06}$&$1.34^{+0.07}_{-0.06}$&226.73/238&$42.76^{+0.03}_{-0.04}$&$-2.42^{+0.03}_{-0.04}$&$43.87^{+0.03}_{-0.04}$\\
      XRT6&60203&--&--&$1.45^{+0.21}_{-0.22}$&$0.21^{+0.04}_{-0.03}$&33.48/30&$42.02^{+0.11}_{-0.10}$&$-3.17^{+0.10}_{-0.11}$&$43.12^{+0.10}_{-0.11}$\\
     XRT7&60459&--&--&$1.47^{+0.11}_{-0.12}$&$0.28^{+0.02}_{-0.03}$&81.18/79&$42.13^{+0.07}_{-0.06}$&$-3.06^{+0.07}_{-0.06}$&$43.23^{+0.07}_{-0.06}$\\
\hline
     \end{tabular}
\leftline{$\dagger$ in the unit of ~\normflux.}
\end{table*}
\subsection{X-ray Spectral Analysis}
\label{3.2}
We carried out X-ray spectral analysis using {\tt XSPEC v12.12.1} \citep{1996ASPC..101...17A}. Spectral analysis was performed using one epoch of \xmm~observation (XMM) and eight epochs of \swift/XRT observations (XRT1--XRT7) in the energy range of 0.3-10 keV. Furthermore, we analyzed data from simultaneous \swift/XRT and \nustar~observations (0.3-60 keV range) for one epoch (XRT4+NU1), as well as quasi-simultaneous \swift~and \nustar~observations (0.3-60 keV range) for another epoch (XRT5b+NU2). The observation details are presented in Table~\ref{tab:obs}. The uncertainties in each spectral parameter were calculated using the {\tt error} command in {\tt XSPEC} and reported at 90\% confidence. In this work, we considered the unabsorbed X-ray luminosity ($L_{X}$) in the 2-10 keV energy range. The 2-10 keV luminosity was calculated from each spectrum using the {\tt clumin} task on the {\tt powerlaw} model. We only considered the luminosity of the primary continuum emission. Once we calculated $L_{X}$, we converted it to the bolometric luminosity ($L_{Bol}$) using the Eddington ratio-dependent bolometric correction factor ($k_{bol}$) from \cite{Gupta2024}. The $k_{bol}$ for the 2-10 keV range is given by the following relation:
$\log k_{\rm bol}=C \times (\log \lambda_{\rm Edd})^2 + B\times \log \lambda_{\rm Edd} + A$. 

Here, the constants are $C=0.054\pm0.034$, $B=0.309\pm0.095$, and $A=1.538\pm0.063$. The Eddington ratio is computed as $\lambda_{\rm Edd}=k_{\rm bol} \times L_{\rm x}/L_{\rm Edd} = L_{\rm bol}/L_{\rm Edd}$. We estimated the Eddington luminosity $L_{\rm Edd}=1.95\times10^{46}$ erg~s$^{-1}$, by using the relation $ L_{\rm Edd}=1.3\times10^{38} (M_{\rm BH}/M_\odot)$. The derived parameters are summarized in Table~\ref{tab:powerlaw_fit}.
\begin{figure*} 
	\centering
     \hspace{-1cm}
   \includegraphics[trim={0.0cm 3.0cm 0.0cm 0},scale=1.2]{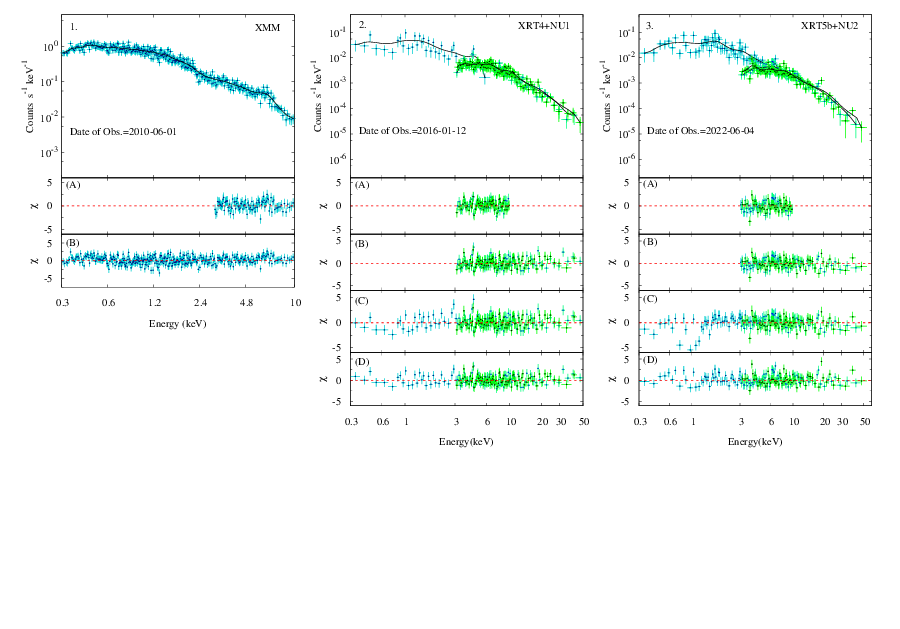}
	\caption{{\it Left panel (1)}: The spectra from June 2010 \xmm~(XMM) observation, fitted with an absorbed power-law model, and panels (A)--(B) represent the corresponding residual plots. {\it Middle and Right panels (2 and 3)}: The broadband spectra from the January 2016 (XRT4+NU1) and June 2022 (XRT5b+NU2) observations, are fitted with the Cutoffpl/CompTT model. Panels (A)--(D) display respective residual plots. Detailed explanations of the figure are provided in Section~\ref{sec:Model}}
	\label{fig:Spectrum} 
\end{figure*}
\subsubsection{X-ray Spectrum of NGC~3822}
\label{sec:Model}
To investigate the spectral variations of the source over an extensive time frame of~$\sim$~16 yr (2008-2024), we constructed a base model covering the broad energy range from 0.3 to 60 keV. The step-by-step model construction is presented in Figure~\ref{fig:Spectrum}. Initially, we consider the spectrum in the 3 to 10 keV range as the X-ray continuum spectrum, and fit it with the power-law model. 
Along with the power law, we also consider Galactic absorption ($N_{H, gal}=3.85\times10^{22} cm^{-2}$) along the line of sight of the source using the multiplicative model {\tt TBabs}~\citep{Wilms2000} in {\tt XSPEC}. Due to low SNR, we applied Cash statistics (C-stat;~\citealt{Cash1979,Kaastra2017}) for spectral fitting. This likelihood-based fitting statistic is used for low-count Poisson data and is particularly effective for bins with a small number of counts. It ensures unbiased parameter estimation in the low-count regime. A {\tt Constant} component is used as a cross-normalization factor while using data from different instruments in simultaneous spectral fitting. The baseline model for the 3-10 keV X-ray spectral fitting is as follows:~{$\tt Constant\times Tbabs\times Powerlaw$} and the corresponding residuals are shown in panel~A of each observation in Figure~\ref{fig:Spectrum}. 

During the primary continuum fitting, we did not observe any significant positive residuals in the iron energy band (6-7 keV) for any of the source spectra (panel A of Figure~\ref{fig:Spectrum}). This indicates that the Fe-line is not detected in these observations; however, Fe could still be present in the disk, but remains undetectable. Next, we extended the spectra to the high-energy regime (above 10 keV) for the broadband observation XRT4+NU1 \& XRT5b+NU2. While extrapolating the primary continuum to the high-energy regime, we did not observe any deviation in the high-energy data points from the primary model for both the broadband observations, as seen in Figure~\ref{fig:Spectrum} (panel B). This suggests that the reflection component in the X-ray spectra of NGC~3822 (above 10 keV) is absent or insignificant during these observation periods. 

Next, we extend the spectra in the lower energy ($<$ 3 keV) range to investigate the presence of any intrinsic absorption in the source spectrum, as well as to examine the presence of a soft X-ray excess. From the power-law fitting, we found that the soft excess component was absent during these observations. To explore the possibility of any intrinsic absorption, we initially used the absorption model {\tt zTbabs} in our baseline model. However, this model is found to be insensitive during spectral fitting and yields the lowest value of the absorption column density. To explore this further, we replaced {\tt zTbabs} with a partially covering absorption model {\tt Pcfabs} to check the presence of any partial absorbers present along the line of sight. While fitting the spectra with the {\tt Pcfabs} model for the observation from 2008 (XRT1) to 2015 (XRT3), we found that no absorption is required; a simple power-law model is sufficient to fit the overall spectra. This result is presented in Figure~\ref{fig:Spectrum} for the {\it XMM-Newton} observation in panel B. However, in the spectra from the 2016 (XRT4+NU1) and 2022 ( XRT5a--XRT5c) observations, a deviation is observed in the low-energy counterpart of the spectrum, as seen in panel C of Figure~\ref{fig:Spectrum}. This indicates the presence of intrinsic absorption components along the line of sight. To account for this, we included an additional absorption component {\tt Pcfabs} in the baseline model of the 2016 (XRT4+NU1) and 2022 ( XRT5a--XRT5c) observations. After 2022, no absorption is required to fit the source spectra. 
\begin{table}
 \setlength{\tabcolsep}{1.5pt} 
\renewcommand{\arraystretch}{1.3} 
    \centering
	\caption{Best-fit parameters for the X-ray broadband observations of NGC~3822 with the model: ${\tt Constant\times Tbabs\times Pcfabs\times Cutoffpl/CompTT}$.} 
	\label{tab:Cutpl+compTT}
	\begin{tabular}{l c r r} 
	\hline
    \hline
    
    Model & Parameter &XRT4+NU1&XRT5b+NU2\\ 
    \hline
    
    {\tt Pcfabs}&$N_{\rm H}(10^{22}cm^{-2})$       &$<0.96$               &$1.44^{+0.75}_{-0.55}$\\
                &$C_{f}$                    &$0.53^{+0.19}_{-0.14}$&$0.69^{+0.08}_{-0.17}$\\
{\tt Cutoffpl}  &$\rm \Gamma$               &$1.68^{+0.04}_{-0.11}$&$1.54^{+0.04}_{-0.22}$\\
                &$\rm E_{cut} (keV)$        &$300(f)$                &$300(f)$  \\
                &$\rm Norm^{\dagger}$           &$1.05^{+0.07}_{-0.21}$&$1.33^{+0.16}_{-0.35}$\\
                &$\rm C-stat/dof$           &191.90/194            &211.70/184\\
    \hline
    \hline
{\tt Pcfabs}    &$N_{\rm H}(10^{22}cm^{-2})$    &$1.59^{+1.84}_{-0.77}$  &$1.47^{+0.45}_{-0.35}$\\
                &$C_{f}$                  &$0.51^{+0.09}_{-0.10}$  &$0.67^{+0.06}_{-0.06}$\\
{\tt CompTT}    &$\rm kT_{e} (keV)$       &$50^{+21}_{-19}$        &$88^{+51}_{-32}$  \\
                &$\rm \tau$               &$0.97^{+0.57}_{-0.34}$  &$0.77^{+0.46}_{-0.31}$\\
                &$\rm Norm^{*}$           &$0.28^{+0.16}_{-0.10}$  &$0.14^{+0.09}_{-0.05}$\\
                &$\rm C-stat/dof$         &192.67/194              &211.11/184\\
   \hline
     \end{tabular}
\leftline{$\dagger$ in the unit of~\normflux.}
 \leftline{* in the unit of~\normflux.}
\end{table}
\begin{figure*} 
	\centering
   \includegraphics[trim={0.4cm 4.0cm 0.3cm 1},scale=1.2]{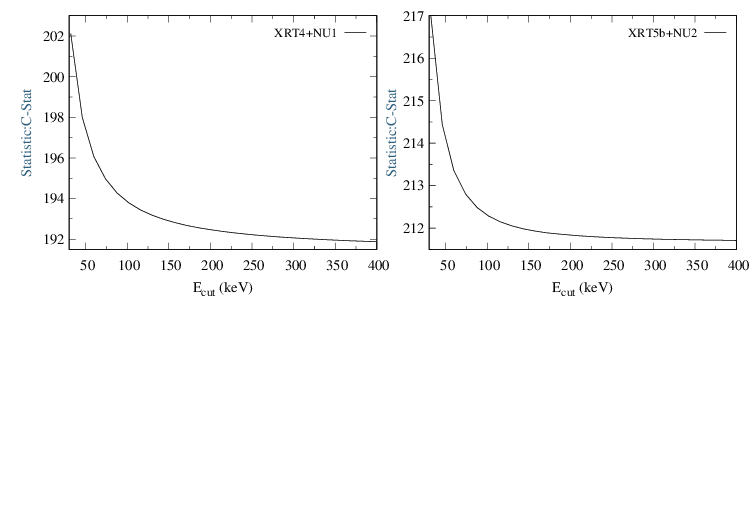}
	\caption{The confidence interval plot of the high-energy cutoff parameter $\rm E_{cut}$ for the observation XRT4+NU1 and XRT5b+NU2. The plots illustrate that $\rm E_{cut}$ could not be constrained owing to low-quality data.}
	\label{fig:Cutpl_contour} 
\end{figure*}
\begin{figure*} 
	\centering
   \includegraphics[trim={0.0cm 0.0cm 0.0cm 0},scale=0.6]{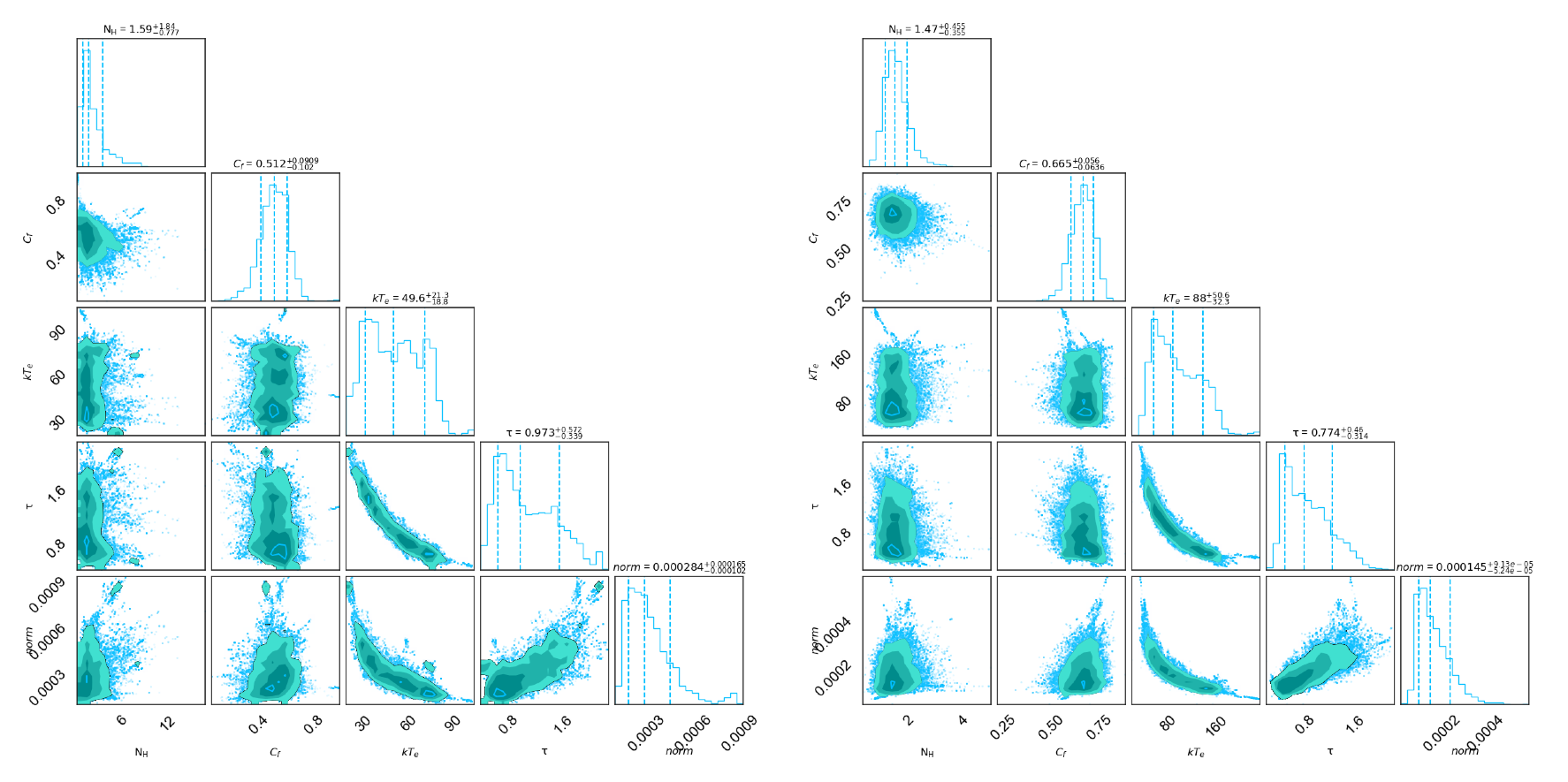}
	\caption{Corner plots of spectral parameters from MCMC analysis for the XRT4+NU1 (left) and XRT5b+NU2 (right) using $\tt CompTT$ model. One-dimensional histograms represent the probability distribution. Three vertical lines in the 1D distribution show 16$\%$, 50$\%$, and 90$\%$ quantiles. We used $\tt CORNER.PY$~\citep{Foreman-Mackey2017} to plot the distributions. The units of $N_{\rm H}$,  $\rm kT_{e}$, and Norm are in $\rm 10^{22}cm^{-2}$, keV, and \normflux, respectively.}
     \label{fig:MCMC}
\end{figure*}

From the spectral fitting in the range of 0.3-10 keV using the baseline model {$\tt Constant\times Tbabs\times Pcfabs \times Powerlaw$}, we estimated that during XRT4+NU1 observation, the hydrogen column density ($N_{H}$)  was $1.20^{+1.01}_{-0.71}\times10^{22}cm^{-2}$ with a covering fraction ($C_{f})$ of $0.62^{+0.09}_{-0.12}$. In observations from XRT5a to XRT5c, the average value of $N_{H}$ was found to be approximately $1.70\times10^{22}cm^{-2}$ with a $C_{f}$ of $0.67^{+0.13}_{-0.23}$. However, no intrinsic absorption is detected in the subsequent observations from XRT5c to XRT7 (see Table~\ref{tab:powerlaw_fit}). The photon index ($\Gamma$) also shows noticeable variability. During the observation period from 2008 to 2024, $\Gamma$ is found to vary between $1.41\pm0.18$ to $1.93\pm0.28$. The fluctuation in $\Gamma$ indicates different spectral states observed during the observational period.

While estimating luminosity in the 2-10 keV range ($L_{X}$), we observed a variation in luminosity across different observations. During the entire duration of observations from 2008 to 2024, the source appeared relatively bright in 2013 (XRT2) and 2022 (XT5a--XRT5d). In 2013, the source luminosity in the 2-10 keV range and the corresponding bolometric luminosity ($L_{Bol}$) are estimated to be $43.14\pm0.08$ and $44.34\pm0.08$, respectively. During this time, the accretion rate (in terms of $\lambda_{Edd}$) is also found to be $-2.06\pm0.08$, the highest compared to other observations. After 2013, the source became fainter, with $L_{X}$ decreasing to $42.35\pm0.10$. However, it brightened again in 2022, reaching an average luminosity of $L_{X}=42.73$. The derived spectral parameters are summarized in Table~\ref{tab:powerlaw_fit}.

After successfully fitting all spectra up to 10 keV, we proceeded to investigate the broadband observations. While fitting the broadband observation, we replaced the power law model with the cutoff power law model to measure the high-energy cutoff ($\rm E_{cut}$) in this AGN. We made the parameter $\rm E_{cut}$ free in all our fits. However, we could not constrain the value of $\rm E_{cut}$ for the broadband observations. To robustly identify the error, we performed some statistical tests, using the {\tt STEPPAR} command in XSPEC. Figure~\ref{fig:Cutpl_contour} illustrates that $\rm E_{cut}$ could not be constrained. We then tested different values for $\rm E_{cut}$ and found that the fit was insensitive to this parameter. As a result, we fixed $E_{cut}$ at 300 keV~\citep{Ueda2014,Ricci2017,Ricci2018} for these observations. The details of the results are presented in Table~\ref{tab:Cutpl+compTT}.  
\begin{figure*} 
	\centering
   \includegraphics[trim={0.0cm 0.0cm 0.0cm 0},scale=0.6]{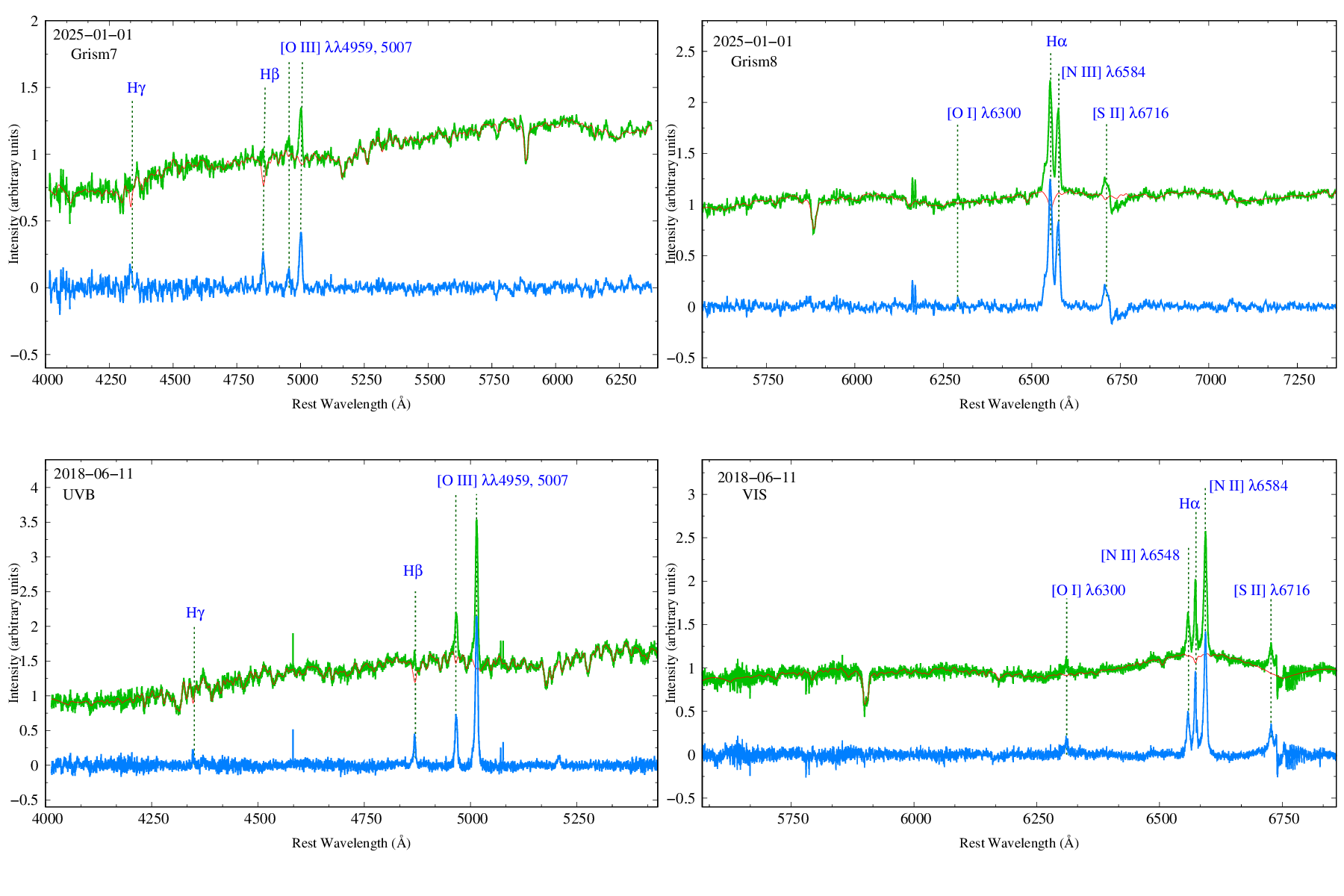}
	\caption{Optical spectrum of NGC~3822 as observed with HFOSC (Gr7 and Gr8) and X-Shooter (UBV and VIS arm). The green line represents the observed spectrum. The red line is the pPXF fit for the stellar component, and the host-galaxy subtracted AGN spectrum is shown in the blue line.}
	\label{fig:ppxf_spec} 
\end{figure*}

To investigate the properties of the Compton cloud, we replaced the cutoff power law model with  {\tt compTT}~\citep{Sunyaev&Titarchuk1980,Titarchuk1994} model. This model provides the electron temperature ($kT_{e}$) and optical depth ($\tau$) of the Compton cloud. During spectral fitting, we fixed the seed photon temperature, $kT_s = 100$ eV, corresponding to the Wien’s temperature of the accretion disc for this source, and adopted a spherical geometry for the Compton cloud. After obtaining good fits with the {\tt compTT} model, we performed an MCMC run on each fit to explore the parameter space. We used a chain length of 100000, 20 walkers, and a burn-in period of 10000 steps.  We then calculated the most likely parameter values and their corresponding errors from the posterior distributions of the chains. Figure~\ref{fig:MCMC} presents the results of the MCMC analysis. From the spectral fitting, we found $kT_{e}$=$50^{+21}_{-19}$ keV for the observation XRT4+NU1. During the outburst phase in the observation XRT5b+NU2, the Compton cloud was hotter, with an electron temperature of $88^{+51}_{-32}$ keV. The optical depth varied across observations within the range of $\sim$0.77--0.97. The results obtained from the spectral fitting are quoted in Table~\ref{tab:Cutpl+compTT}.

\subsection{Optical Spectroscopic Analysis}
We carried out optical spectroscopic analysis using nine epochs of observations between 2018 and 2025 with X-Shooter/VLT and HFOSC/HCT (see Table~\ref{tab:obs}). In order to investigate the properties of the emission lines accurately, we modelled the stellar population of the host galaxy using the Penalised Pixel-Fitting (pPXF) code \citep{Cappellari2023}. This package extracts the stellar population from the absorption-line spectra of galaxies, using a maximum Penalized likelihood approach. We used the MILES \citep{2011A&A...532A..95F} stellar template library in the pPXF fitting. This library features 
$\sim$1,000 stars, with spectra obtained by the Isaac Newton Telescope, and covers the wavelength range of 3525~\AA~ -- 7500~\AA~ at a 2.5~\AA~FWHM resolution. The pPXF method is particularly useful for correcting the emission line profiles for the underlying absorption features. The best-fit stellar model was subtracted from each observed spectrum, leading to a pure AGN spectrum. We detect significant host-galaxy stellar absorption in the ~\hbeta~ line in both Gr7 and UVB spectra (see Figure~\ref{fig:ppxf_spec}). The ~\hbeta~ line is only clearly detected after subtracting the contribution of the host galaxy.  Figure~\ref{fig:ppxf_spec} represents the host-galaxy subtracted and redshift-corrected normalized AGN spectra for a set of observations.

The optical spectra of NGC~3822 reveal several prominent permitted emission lines such as Balmer emission lines \halpha, \hbeta, as well as narrow forbidden emission lines such as $\OIII$, $\OI$, $\NII$, and $\SII$.

\subsubsection{Emission line profiles and their evolution}
\label{sec:FWHM_FWZI}
\begin{figure*} 
	\centering
    \hspace{-1cm}
    \includegraphics[trim={0.0cm 0cm 1cm 0},scale=1.4]{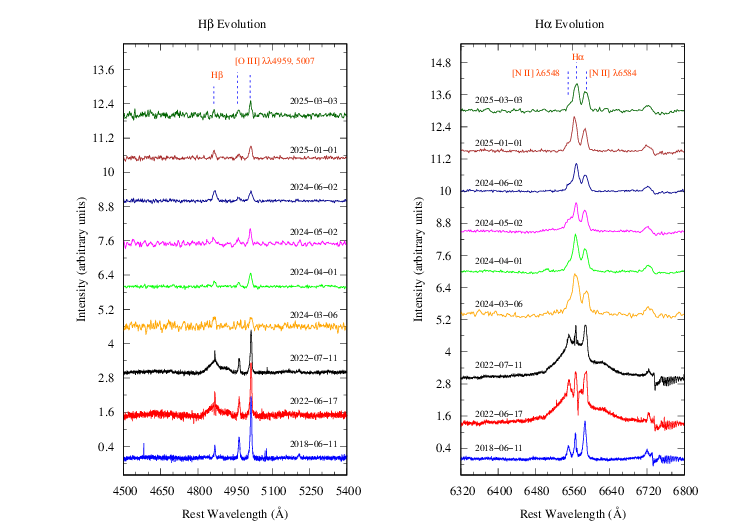}
	\caption{Evolution of the line profiles in \hbeta~and \halpha~regions. All the spectra have been corrected by subtracting the host galaxy contribution. Observation dates are annotated on each spectrum for reference. For clarity, the spectra are plotted with vertical offsets.}
	\label{fig:Hlpha_multiplot} 
\end{figure*}

Optical spectroscopic monitoring of NGC~3822 reveals variation in emission lines over time. To examine the evolution and shape of the \halpha~and \hbeta~line profiles, we focused on plotting only the portion of the optical spectra that covers the \halpha~and \hbeta~emission line regions. The evolution of the emission line profiles across all optical observations is shown in Figure~\ref{fig:Hlpha_multiplot}. The figure indicates that the spectrum of the June 2018 (XS1) observation exhibits a narrow \hbeta~component, while a broad \hbeta~line appears in the 2022 (XS2 $\&$ XS3) observation. During this one-year interval, the evolution of the \hbeta~line profile was also observed (see Figure~\ref{fig:Hlpha_multiplot}). During the HCT1 to HCT3 observations, the broad \hbeta~line is not very prominent. However, after a three-month observational gap, in June 2024 (HCT4), a prominent \hbeta~line was observed. Following a six-month gap, in January 2025, the \hbeta~line was again detected and became very weak after two months in March 2025 (HCT6), which is similar to what was observed in March 2024 (HCT1). The narrow forbidden $\OIII$ lines are detected in all epochs. The broadness of the \hbeta~line significantly decreased during the monitoring period from 2024 (HCT1) to 2025 (HCT6). In the 2018 observation, visual inspection of the \halpha~region, reveals that the broad component of \halpha~is either absent or significantly weakened, while the narrow components of \halpha~and the $\NII$ lines on either side of \halpha~are clearly visible. In contrast, the 2022 observations show the unambiguous emergence of the broad \halpha~component. Superimposed on the broad \halpha~profile are the narrow components of \halpha~and $\NII$. Post 2022, there is gradual decline in the strength of the broad \halpha~component, as well as in the narrow \halpha~and $\NII$ lines. The profile changes noticed in the broad \halpha~line (see Fig.~\ref {fig:Hlpha_multiplot}; right panel) are also seen in the broad component of \hbeta~(see Fig.~\ref {fig:Hlpha_multiplot}; left panels)

\begin{figure*}
     \centering
    \hspace{-0.05in}
         \includegraphics[trim={1 0cm 0cm 0},clip,width=0.82\textwidth]{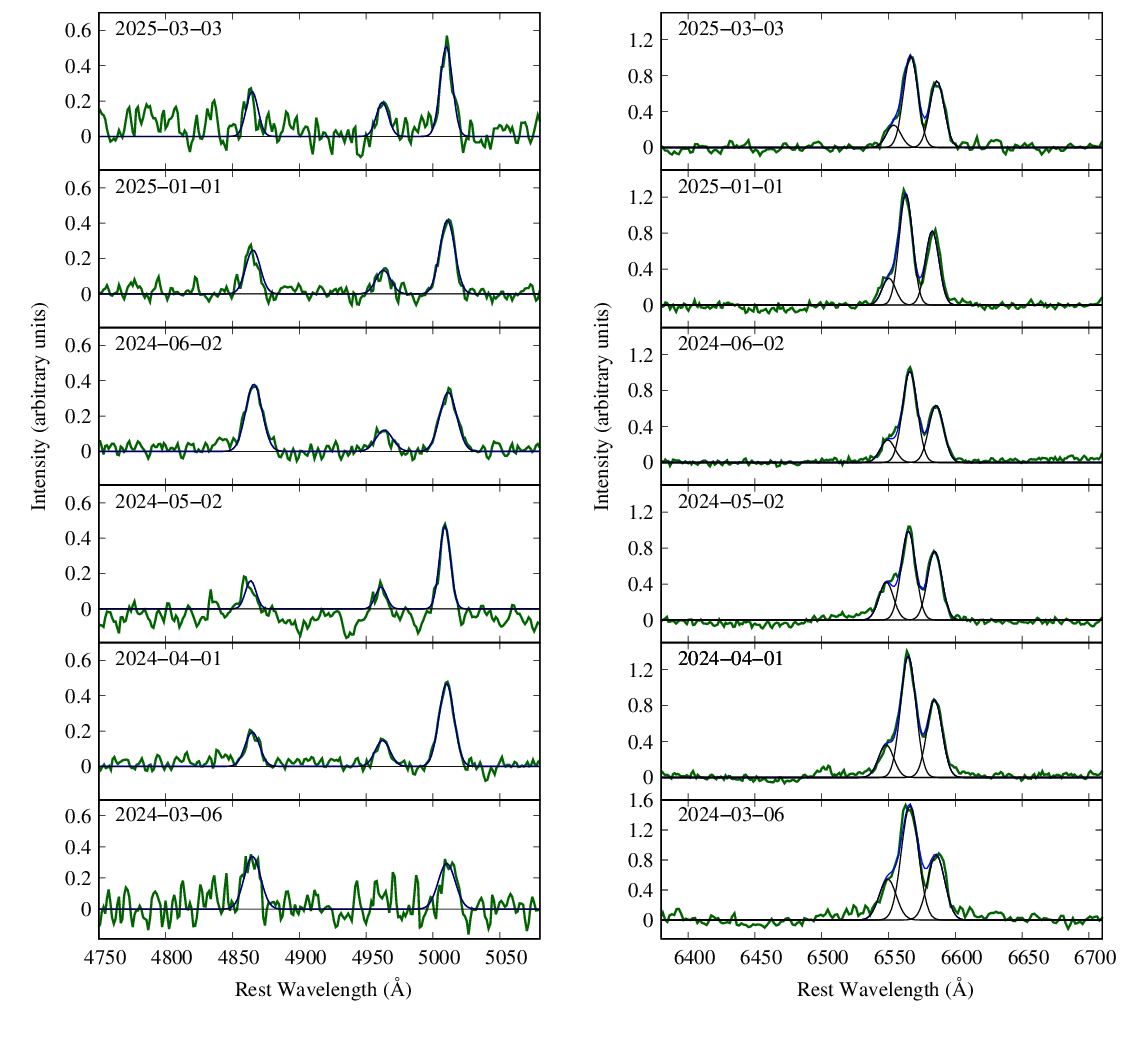}
     \hfill
     \hspace{-0.20in}
         \includegraphics[trim={1 8cm 0cm 0},clip,width=0.82\textwidth]{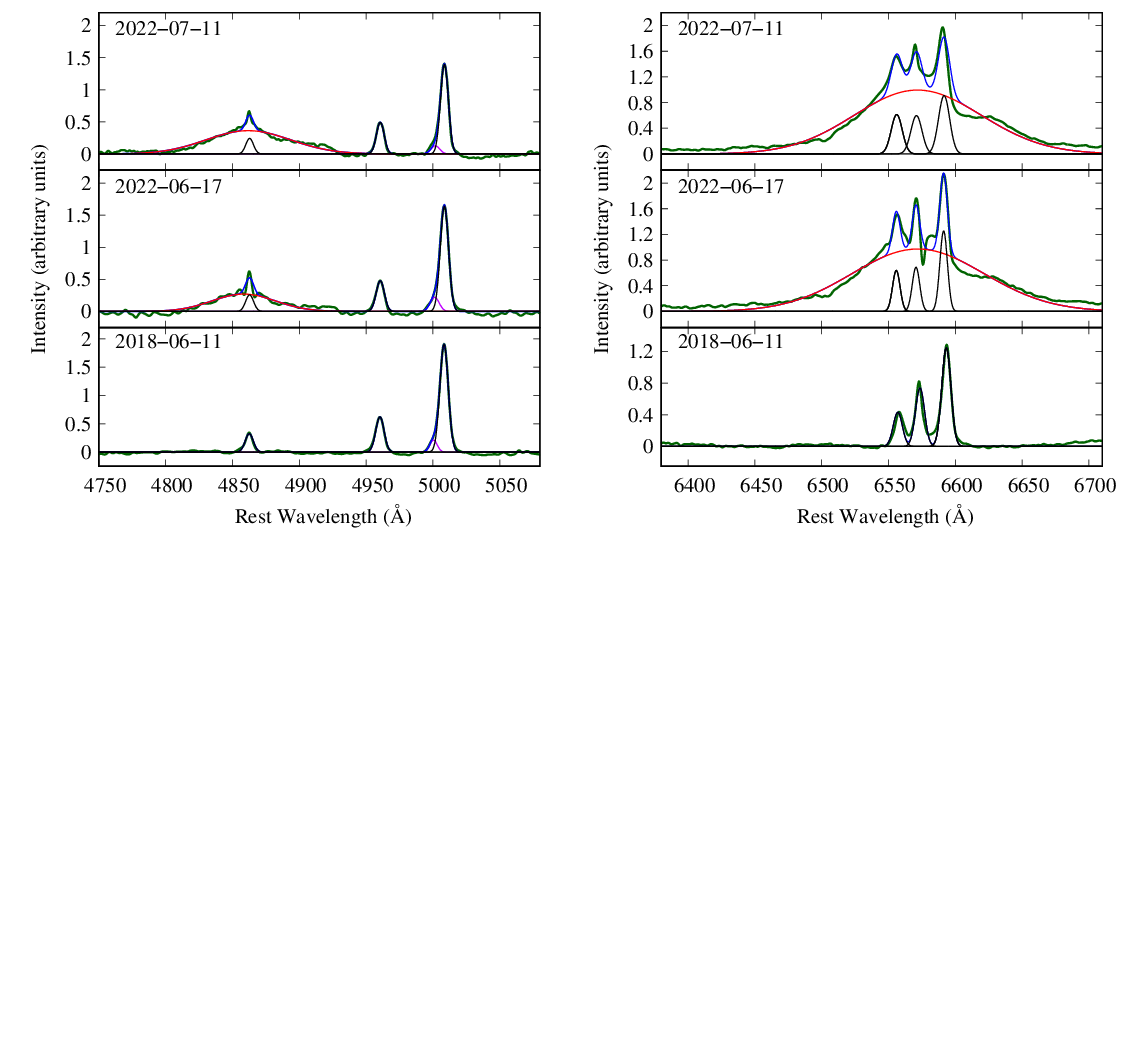}
         
     \caption{The \hbeta~ and $\OIII$ (left panels) as well as \halpha+[NII] regions (right panels) are shown for all the epochs of HCT (top six panels) and X-Shooter (bottom three panels) observations of NGC~3822. For both the spectral ranges, the green lines represent the observed spectra, the blue line is the fitted model, and the black line represents each narrow emission line. A clear Balmer broad component (in red) appears in the 2022 spectra. }

        \label{fig:Hbeta_multiplot}
\end{figure*}

To further investigate the line profile shapes, we calculated the full width at half maximum (FWHM) of the emission lines. We focused on the most prominent spectral lines, which include the \halpha+$\rm {[N\,II]}$ complex, \hbeta~and the $\OIII$. We fit the emission lines by the multicomponent spectral fitting code {\tt pyQSOFIT}~\citep{Guo2018}. The narrow components of \halpha~, $\NII$, \hbeta~ and $\OIII$ are modelled with narrow Gaussian components. As these narrow emission lines are known to originate from the same NLR, the width of these lines should be comparable (within errors). For that reason, the line widths and velocity offsets of the narrow lines were tied to each other. During the spectral fitting, for broad \halpha~and \hbeta~, broad Gaussian profiles are used. The X-Shooter offers higher spectral resolution than the HFOSC. Therefore, we applied an instrumental resolution correction before calculating the FWHM of the emission lines. The calculated values of the line width are given in Table~\ref{tab:FWHM} and the Gaussian profile fitted plots are illustrated in Figure~\ref{fig:Hbeta_multiplot}.

In the 2018 optical spectrum, the $\OIII$ emission lines appeared narrow, with a FWHM of approximately 444~\kms. In subsequent observations between 2022 and 2025, the FWHM did not change significantly, exhibiting a maximum FWHM of less than 900\kms. However, a dramatic change in line profile was seen in the \hbeta~region during the long-term optical monitoring period from 2018 to 2025. In the 2018 observation, \hbeta~line appeared only as a narrow component with a FWHM of $444\pm107$\kms. By 2022, this narrow component was still present, but the profile showed a clear broadening, due to which an additional broad ($\rm \hbeta$) component was required to fit the spectrum along with the narrow \hbeta~ component. The FWHM of the $\rm \hbeta$ line was found to be $3454\pm795$ and $4578\pm1054$ \kms~during the June 2022 and July 2022 observations, respectively. The FWHM of the narrow \hbeta~ component during these observations was found to be $<500$ \kms.  In later observations during 2024-2025, the broad \hbeta~ component disappeared, though the narrow component remained in the spectrum with FWHM $<1000$ \kms.

A similar evolution of the line profiles was also seen in the \halpha~region. In the 2018 spectrum, the line profile was narrow and well fitted with three Gaussian components representing narrow \halpha~and $\NII$ doublet. However, both 2022 observations required an additional broad \halpha~component to model the observed profile, along with the narrow \halpha~and $\NII$ lines. The broad \halpha~component exhibited FWHM values of $5311\pm1222$ and $5110\pm1176$ \kms~during the June and July 2022 observations, respectively, while the FWHM of narrow \halpha~ and $\NII$ doublet did not show any notable changes (Table~\ref{tab:FWHM}). During the 2024-2025 monitoring period, the broad \halpha~component was no longer required and the emission lines were adequately fitted with only the narrow components (Figure~\ref{fig:Hbeta_multiplot}). In this time frame, the FWHM of the narrow \halpha~line varied between 543 and 684 \kms. The FWHM values are tabulated in Table~\ref{tab:FWHM}. While fitting the [O III]$\lambda5007$ line in the X-Shooter data, we noticed a hump like feature in the bluer wing. To model the feature, we added a Gaussian component. The FWHM was estimated to be $391\pm90$, $480\pm110$, and $461\pm106$\kms, during the June 2018, June 2022, and July 2022 X-Shooter observations, respectively. This suggests the presence of a possible weak outflow signature in the source. However, this feature was not detected in any of the HCT observations, likely due to poor energy resolution of the instrument and also poor SNR during the observations.

\begin{table*}
\centering
\caption{FWHM values (in km s$^{-1}$) for different emission lines observed between June 2018 and March 2025.}
\label{tab:FWHM}
\scriptsize
\noindent\hspace*{-3.0cm}%
\setlength{\tabcolsep}{2.2pt}
\resizebox{1.2\textwidth}{!}{
\begin{tabular}{lcccccccc}
\hline
Date & $\hbeta$ & $\hbeta$ & [O III]$\lambda4959$ & [O III]$\lambda5007$ & [N II]$\lambda6548$ & $\halpha$ & $\halpha$ & [N II]$\lambda6584$ \\
     &(Narrow) &(Broad)            &                      &                      &                     &(Narrow) &(Broad)& \\
\hline
2018-06-11 &$444\pm107$  & -- &$444\pm107$&$444\pm107$&$367\pm84$ &$367\pm84$ & -- &$367\pm84$  \\
2022-06-17 &$435\pm102$  &$3454\pm795$ &$435\pm102$&$435\pm102$&$360\pm70$&$360\pm70$& $5311\pm1222$ &$360\pm70$ \\
2022-07-11 & $446\pm102$ &$4578\pm1054$ &$446\pm102$&$446\pm102$&$447\pm102$ &$447\pm102$& $5110\pm1176$&$447\pm102$ \\
2024-03-06 & $893\pm205$ & -- & -- &$893\pm205$ &$684\pm157$ &$684\pm157$ & -- &$684\pm157$ \\
2024-04-01 & $790\pm182$ & -- & $790\pm182$ & $790\pm182$ & $613\pm141$ & $613\pm141$ & -- & $613\pm141$ \\
2024-05-02 & $582\pm134$ & -- & $582\pm134$ & $582\pm134$ & $585\pm134$ & $585\pm134$ & -- & $585\pm134$ \\
2024-06-02 & $814\pm205$ & -- & $814\pm205$ & $814\pm205$ & $595\pm136$ & $595\pm136$ & -- & $595\pm136$ \\
2025-01-01 & $790\pm182$ & -- & $790\pm182$ & $790\pm182$ & $544\pm125$ & $543\pm125$ & -- & $544\pm125$ \\
2025-03-03 & $646\pm149$ & -- & $646\pm149$ & $646\pm149$ & $580 \pm 133$ & $580 \pm 133$ & -- & $580 \pm 133$ \\
\hline
\end{tabular}
}
\end{table*}

\section{Discussion}
\label{sec:discussion}
We conducted a multiwavelength study of NGC~3822 using the data from \xmm, \swift, and \nustar, HCT, and VLT observatories. 
Our optical monitoring reveals the appearance and disappearance of broad emission lines, indicating the changing-look nature of NGC~3822. Here, we discuss the long-term behavior of the source in light of multiwavelength observations. 

\subsection{Outburst in NGC~3822}
From the long-term multiwavelength light curve (left panel of Figure~\ref{fig:all_flux}), it is evident that the source exhibited significant flux variability across all observed bands during the period from 2008 to 2024. In 2013, a notable increase in flux is observed, particularly in the X-ray and UVW1 bands, where the X-ray flux increased by a factor of $\sim$4, while the UVW1 flux increased by a factor of $\sim6$. After a nine-year observation gap, the source again showed a significant enhancement of flux across all bands during the observations in 2022. The X-ray flux increased to an even higher intensity level by a factor of $\sim$5 until December 2022 (Figure~\ref{fig:all_flux}). After that, the X-ray flux showed a dramatic decrease by a factor of $\sim$16 during the 2023-2024 observation period. The right panel of Figure~\ref{fig:all_flux} presents the detailed multiwavelength light curve during the 2022 observation period. It is found that the increasing and decreasing trends of flux are significant in X-rays and UV, while it is less pronounced in the optical U, V, and B bands, possibly due to contamination from the host galaxy. This sudden rise in flux in different wavebands indicates a nuclear outburst in NGC~3822. The discovery of an outburst in the nucleus of NGC~3822 was previously reported by \cite{Hinkle2022}, and subsequent optical UVB photometry conducted by \cite{Oknyansky2022} reported that the outburst persisted in NGC~3822 between 21 March and 24 April 2022.

Long-term monitoring of the source with \swift, spanning from 2008 to 2024, reveals a flux enhancement in 2013 (Figure~\ref{fig:all_flux}). However, due to limited observations around 2013, no definitive conclusions can be drawn regarding this flux increase. After approximately 9 years, in 2022, the source again exhibited a significant flux increase, observed across multiple wavebands (from optical to X-ray) during March and April, which confirms the nuclear outburst in NGC~3822 as reported by \cite{Hinkle2022}.

\subsubsection{Possible cause of the outburst}
An outburst in AGN can, in principle, be caused by a tidal disruption event (TDE), or sudden changes in the accretion process. The TDE is characterized by a dramatic increase in luminosity, showing single sharp outbursts in X-rays ~\citep{Lin_2011} or optical–UV bands \citep {vanVelzen2021} which then fades away on a timescale of months to years \citep{Rees1988,Komossa2002}. In general, for a classic TDE, after the star is tidally disrupted by an SMBH, the luminosity is expected to decay following a $t^{-5/3}$ trend over a timescale of a few hundred days\citep{Rees1988,Komossa2015,Komossa2017,Gezari2021}. However, multiwavelength studies indicate systematic differences in the fallback emission across different wavebands. In the optical/UV band, the fallback decay is expected to be slower, following the $t^{-5/12}$ trend \citep{Lodato2011}. Although some observations suggest that the optical/UV light curves broadly follow the $t^{-5/3}$ decay, several well-observed events have shown significant flattening \citep{Gezari2021} or even plateau-like behavior($\approx t^{0}$)~\citep{Kajava2020}. 

\begin{figure} 
	\centering
	\includegraphics[trim={1cm 4.5cm 2cm 0},scale=1.1]{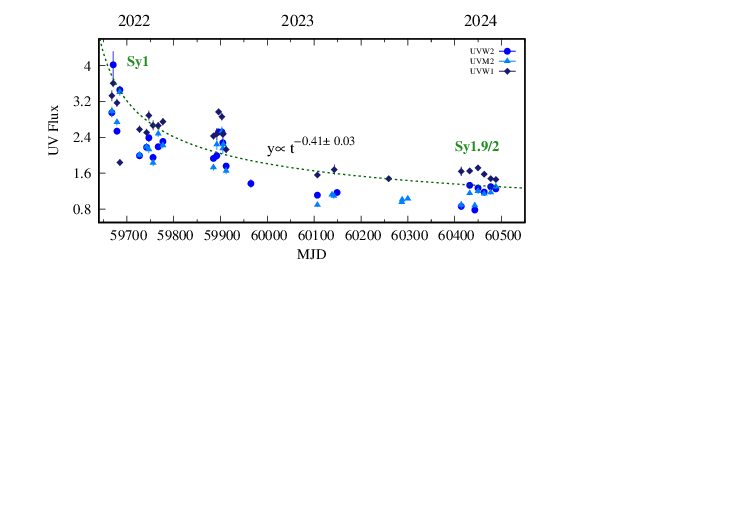}
    \caption{The UV flux variation observed during the outburst phase in 2022 to the post-outburst phase from 2023 to 2024. The fitted black dashed line indicates the decay profile.}
	\label{fig:TDE} 
\end{figure}

To investigate the potential role of the TDE as a cause of the outburst observed in 2022 in NGC~3822, we fitted a decay profile to the UV flux variation observed during the outburst phase in 2022 through the post-outburst phase that spanned from 2023 to 2024. We observed that after the initial ﬂare in 2022, the UV flux monotonically dimmed with $t^{-0.41\pm0.03}\approx t^ {-5/12}$ as shown by the dashed line in Figure~\ref{fig:TDE}. This slope is consistent with the theoretical prediction of \citet{Lodato2011}. The $t^{-5/12}$ decay arises from the thermal emission of a cooling accretion disc formed after the tidal disruption event. Following the disruption, the accretion rate is also expected to decay. We examined the evolution of the accretion rate from the 2022 to 2024 observations along with the evolution of the UV flux. We found that during the outburst the accretion rate was $\lambda_{Edd}\sim3.8\times10^{-3}$. However, in the decay phase $\lambda_{Edd}$ decreased to $\sim0.7\times10^{-3}$, approximately five times lower than the value observed during the 2022 outburst.

Optical spectroscopic studies suggest that TDE can produce temporary emission features, such as strong broad Balmer emission lines that may appear double-peaked (e.g.~\citealt{Komossa2008,Holoien2014,vanVelzen2021}). The optical spectra of NGC~3822 were taken with the X-Shooter instrument from the VLT in June-July 2022, approximately two months after the outburst reported by \cite{Oknyansky2022}. The effect of the outburst is clearly observed in the optical spectra (Figure~\ref{fig:Hlpha_multiplot}). We found evidence of changes in the line profiles, accompanied by the appearance of the broad \halpha~and \hbeta~emission lines, where the broad~\halpha~region is blended with narrow forbidden $\NII$ lines (Figure~\ref{fig:Hlpha_multiplot}). However, in the observations after 2022, this complexity in the emission lines was no longer present. Similar TDE-like spectral features have also been observed in the optical spectrum of other AGNs, such as SDSS~J0952+2143~\citep{Komossa2008} and SDSS~J1617+0638~\citep{Zhang2024}. The classification of TDEs can be found in the recent publication by \cite{vanVelzen2021}. In this study, TDEs that give rise to broad emission features of H$\alpha$ and H$\beta$ in the optical spectra of AGNs are classified as TDE-H, the most common subclass among all TDEs. During the outburst, the optical spectra of NGC~3822 displayed such broad emission line (BEL) features. Considering the optical spectral features observed in 2022, along with the decay trend in the UV flux, we conclude that the sudden outburst observed on the NGC~3822 is likely associated with a tidal disruption event.

\subsection{Long-term and short-term variability in flux }
In order to quantify the variability strength in AGN, we calculated the $\rm F_{var}$ across the X-ray, UV, and optical bands for the long-term (2008-2024) and short-term (2022) monitoring campaign (Section~\ref{sec:Fvar}). The results are summarized in Table~\ref{tab:F_var}. Our analysis reveals that $\rm F_{var}$ is stronger at shorter wavelengths. This anticorrelation suggests that the strength of the fractional variability can be considered as an indicator of the distance of the accretion disk emission with respect to the central source~\citep{Zetzl2018}. This trend is consistent with previous studies of other AGNs, such as HE~1136-2304~\citep{Zetzl2018}, NGC~5548~\citep{Fausnaugh2016}, Mrk~509 \citep{Kumari2021}, and IRAS~23226-3843~\citep{Kollatschny2020,Kollatschny2023}. Furthermore, we modeled the wavelength dependency of the variability using a power-law model: $\rm F_{var}=a\times\lambda^{-c}$ as indicated by the dotted line in Figure~\ref{fig:Fvar}. The optimal value we obtained for c from the long-term observation is $\sim$2.51, indicating a steeper decline in variability. However, the short-term monitoring campaign in 2022 revealed a less steep decline, with the value of $c$ decreasing to about 1.91. This value is also comparable to those found for other CLAGN, e.g., IRAS~23226-3843 \citep{Kollatschny2023}, reported a value of $\rm c\sim2.89$ for their long-term monitoring. However, an extensive variability campaign conducted on NGC~5548 in 2014 \citep{Fausnaugh2016} yielded a value of c=0.74 based on the power-law fit.  
Figure~\ref{fig:Fvar} also indicates that the observed X-ray emission does not exactly follow the same trend as UV / optical emission, suggesting that the origin of the X-ray and UV / optical photons is different. A similar behavior has been found in the changing-look AGN HE~1136-2304~\citep{Zetzl2018}. This anti-correlation refers to the fact that the UV/optical continuum emission is generally associated with the emission from accretion disk~\citep{Shakura1973,1995ApJ...455..623C,Hubeny2001}, while the X-ray emission is linked to the process of Comptonization of disk photons in a hot electron cloud or corona. The variability pattern in NGC~3822 suggests significant flux variations from the X-ray to UV bands, but less notable variability in the optical band, possibly due to contamination by the host galaxy in the optical band.
\subsection{X-ray Spectral Properties of NGC~3822.}
For the first time, we investigated the spectral properties of NGC~3822 using long-term X-ray observations from 2008 to 2024. 
Throughout our observations from 2008 and 2024, we noticed a clear variation in the photon index of the primary continuum. The photon index varies between $1.41\pm0.18$ and $1.93\pm0.28$. In the 2008 to 2013 observation, we observed a similar type of continuum with $\Gamma\sim1.54$, indicating a relatively harder spectrum. During the 2015 XRT3 observation, $\Gamma$ further decreased to $1.41\pm0.18$. However, the spectrum was soft during the 2016 (XRT4 + NU1) to 2022 (XRT5c) observations, with an average $\Gamma \sim 1.77$. After the XRT5c observation, the spectrum transitioned back to a relatively harder state compared to the previous observation. Such spectral transitions have also been identified in other AGNs, e.g., NGC~1365 \citep{Liu2021}, Mrk~6 \citep{Layek2024}, NGC~1566 \citep{Titarchuk2025}, UGC~6728 \citep{Nandi2024}.

Along with the variation in the photon index, the source also exhibited a variation in the accretion rate during 2008--2024. We investigated possible correlations between the photon index ($\Gamma$) and the accretion rate ($\log\lambda_{Edd}$). We utilized the Pearson Correlation Coefficient (PCC)\footnote{\url{https://www.socscistatistics.com/tests/pearson/default2.aspx}} to check the order of correlations between the spectral parameters.
We found a weak positive correlation between $\Gamma$ and $\log\lambda_{Edd}$. The $\Gamma$-$\lambda_{Edd}$ correlation suggests a possible connection between the accretion disk and the hot X-ray corona \citep{Vasudevan2007, Fabian2015, Kawamuro2016, Ricci2018, Layek2024, Layek2025}. This correlation is typically more pronounced in sources at high accretion rates ($\lambda_{Edd}>0.3$ ), where steep photon indices ($\Gamma>2$) are commonly observed \citep{Shemmer2008,Risaliti2009,Brightman2013}. However, in the case of NGC~3822, we found a low accretion rate of $\lambda_{Edd}\sim10^{-3}$ with the photon index in the range $\Gamma\sim1.4-1.9$, indicating only a weak correlation between these parameters.

\begin{figure} 
	\centering
	\includegraphics[trim={0cm 1.0cm 1cm 0},scale=0.8]{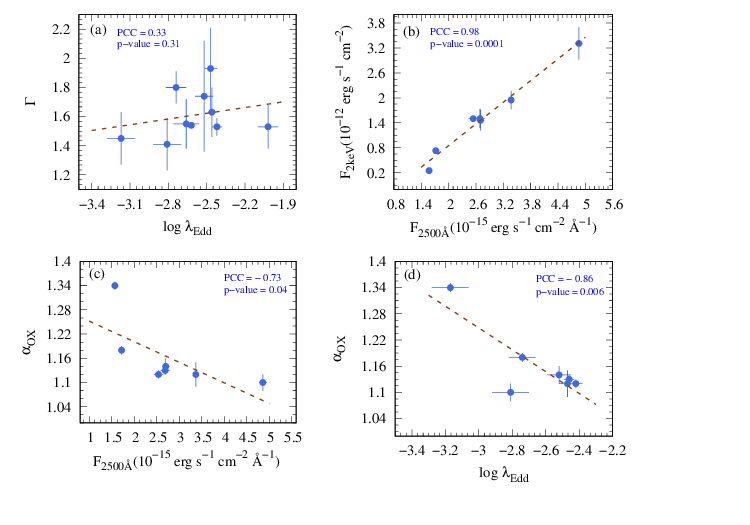}
    \caption{Correlation between different spectral parameters.}
	\label{fig:Correlation} 
\end{figure}

Another important diagnostic parameter to understand the connection between the cold accretion disk and the hot X-ray emitting corona is the $\alpha_{\rm ox}$. This parameter represents the slope of a hypothetical power-law between the UV and X-ray regimes. The parameter is defined as: $\alpha_{\rm ox}=-0.384\rm log(F_{\rm 2keV}/F_{\rm 2500\text{\AA}}$) \citep{Tananbaum1979,Strateva2005,Lusso2010}. We used the UVW1 filter to compute $\alpha_{\rm ox}$ as it is the one with the closest effective wavelength to 2500~\AA. The flux values and the corresponding $\alpha_{\rm ox}$ are summarized in Table~\ref{tab:alphaOX}. We first examined the well-established correlation between the X-ray flux at 2 keV ($F_{\rm 2keV}$) and the UV flux at 2500~\AA~($F_{2500\text{\AA}}$). This relation between UV and X-ray is important to test the energy generation mechanism in AGN. We found a strong positive correlation between $F_{\rm 2keV}$ and $F_{2500\text{\AA}}$ with a PCC value of $0.98$ and a corresponding p-value of $10^{-4}$. The observed correlation supports the disk–corona interaction scenario, where soft UV photons from the accretion disk are up-scattered to higher energies (X-rays) by the relativistic electrons present in the hot corona through inverse Compton scattering (e.g. \citealt{1991ApJ...380L..51H,Haardt1993H}). As this process continues, the corona gradually cools, via inverse Compton scattering, making it less efficient at producing X-rays when the disk becomes more luminous. Consequently, an anti-correlation between $\alpha_{\rm ox}$ and $F_{2500\text{\AA}}$ is expected. In our investigation on NGC 3822, we found that $\alpha_{\rm ox}$ is anti-correlated with UV luminosity with PCC=$-$0.73 with p-value=0.04. This trend is consistent with the previous studies that reported a significant anti-correlation between $\alpha_{\rm ox}$ and $F_{\rm 2500\text{\AA}}$ (e.g.,~\citealt{Strateva2005,Vasudevan2009,Lusso2010,Lusso2012}). Further, we checked $\lambda_{Edd}$-$\alpha_{\rm ox}$ correlation. We found a strong anti-correlation between these two parameters. The dependence of $\alpha_{\rm ox}$ with $\lambda_{\rm Edd}$ suggests that the ratio between the X-ray and UV flux decreases with increasing Eddington ratio, implying that increased accretion leads to weaker coronal emission \citep{Lusso2010,Fanali2013,Layek2025}. The correlation plots along with the correlation coefficient values are shown in Figure~\ref{fig:Correlation}.
\begin{table}
 \setlength{\tabcolsep}{1.6pt} 
 \renewcommand{\arraystretch}{1.5} 
    \centering
	\caption{Observed value for the optical/UV to X-ray flux ratio parameter $\alpha_{\rm ox}$.} 
	\label{tab:alphaOX}
	\begin{tabular}{lcccc} 
	\hline
     \hline
     ID  & MJD    &$F_{\rm 2keV}$    &UVW1         &$\alpha_{\rm ox}$\\
         &        &($10^{-12}$)  &($10^{-15}$)     & \\
    \hline
XRT2      &56499  &$3.31\pm0.39$& $4.86\pm 0.08$& $1.10\pm 0.02$\\
XRT4+NU1  &57339  &$0.73\pm0.02$& $1.72\pm 0.07$& $1.18\pm 0.01$\\
XRT5a     &59697  &$1.95\pm0.22$& $3.37\pm 0.05$& $1.12\pm 0.03$\\
XRT5b+NU2 &59745  &$1.50\pm0.23$& $2.69\pm 0.10$& $1.13\pm 0.01$\\
XRT5c     &59772  &$1.46\pm0.25$& $2.70\pm 0.08$& $1.14\pm 0.02$\\
XRT5d     &59898  &$1.50\pm0.02$& $2.54\pm 0.10$& $1.12\pm 0.01$\\
XRT6      &60203  &$0.25\pm0.02$& $1.57\pm 0.06$& $1.34\pm 0.01$\\
XRT7      &60459  &$0.30\pm0.01$& $1.60\pm 0.05$& $1.31\pm 0.02$\\

\hline
\end{tabular}
\begin{minipage}{08cm}
\small \textbf{Notes:} The $F_{\rm2keV}$ represents the flux in the 1.5--2.5 keV energy band. The UVW1 monochromatic flux is obtained by averaging the fluxes from individual observations within each segment. The X-ray flux is in the unit of \flux and UV flux is in the unit of \efluxA.
\end{minipage}
\end{table} 
\subsubsection{Coronal properties}
In this work, we used two broadband observations, XRT4+NU1 and XRT5b+NU2, covering the energy range of 0.3-60 keV, to investigate the high-energy properties of this AGN. Initially, we fitted the primary X-ray continuum in the 2-10 keV range with a simple power-law model. While extending the primary continuum up to 60 keV, we did not observe any deviation in the spectra (see Figure~\ref{fig:Spectrum}), indicating the absence of a reflection hump above 10 keV. However, the availability of hard X-ray data beyond 10 keV provided us with an opportunity to probe the coronal properties of the AGN. We begin with an absorbed cutoff power law model that contains two primary spectral parameters that carry information on the physical properties of the X-ray corona, e.g., $\Gamma$ and the cutoff energy ($E_{c}$). However, because of the low signal-to-noise ratio in the data, we were unable to constrain the cutoff energy. Next, we adopted the widely used Comptonization model, {\tt compTT} \citep{Titarchuk1994}, which can provide insights into the physical parameters of the corona, such as its temperature ($kT_{e}$) and optical depth ($\tau$). The coronal parameters derived from the spectral fit are listed in Table~\ref{tab:Cutpl+compTT}.  From the 2016 observation, we obtained the coronal temperature of $kT_{e}=50^{+21}_{-19}$ keV and an optical depth of $\tau=0.97^{+0.57}_{-0.34}$.  After a six-year gap, during the 2022 observations, we found that the electron temperature of the corona is $88^{+51}_{-32}$ keV with an optical depth of $0.77^{+0.46}_{-0.31}$. The lower optical depth value ($\tau < 1$) indicates an optically thin plasma. From both epochs, we found a significant degeneracy between coronal temperature and optical depth, as illustrated in Figure~\ref{fig:MCMC}. This type of anticorrelation between $kT_{e}$ and $\tau$ has been detected in previous studies of AGNs in hard X-rays with NuSTAR \citep{Tortosa2018, Nandi2021, Kamraj2022, Serafinelli2024}.
\begin{figure*} 
	\centering
	\includegraphics[trim={2cm 2cm 2cm 2},scale=1.2]{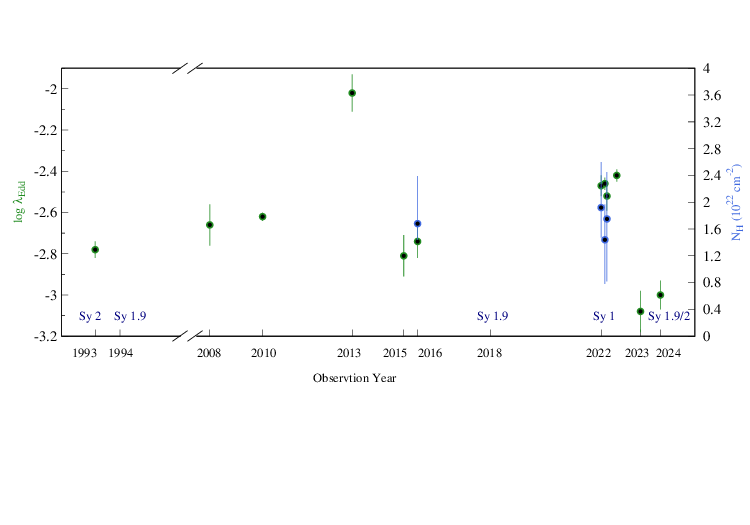}
    \caption{Temporal evolution of the Eddington ratio ($\rm log\lambda_{Edd}$) (black) and the line of sight hydrogen column density ($N_{\rm H}$) (blue) for different observational epochs. The blue text along the axis (e.g., Sy~2, Sy~1.9, Sy~1) indicates the spectral classification of the source at specific epochs. The breaks in the X-axis represent observational gaps.}
	\label{fig:Spectral_type} 
\end{figure*}

\subsection{Variation of the absorber}
We also investigated the presence of any intrinsic absorber in NGC~3822 over this long-term observational period. During 2008--2015 observations, we found no evidence of intrinsic absorption in this source (see Section~\ref{sec:Model}). However, during the 2016--2022 observation period, the presence of an absorber is apparent. In these observations, when we attempted to fit the observed spectra using a simple power law model, we observed a deviation from the power-law continuum below 2 keV (see Figure~\ref{fig:Spectrum}), indicating the presence of a potential absorber. This absorption feature is best described by a partially covering absorber model. The partially covering absorber is detected from the 2016 observation up to the XRT5c observation in 2022. During this period, the line-of-sight column density ($N_{H}$) and the corresponding covering fraction show a consistent presence of intrinsic absorption. However, no significant absorption is detected in the data from XRT5d and later observations. Based on long-term observations from 2008 to 2024, it is found that NGC~3822 exhibits a systematic change in column density. Fitting the spectra with the partially covering absorber model clearly indicates that the material around the SMBH is clumpy in nature. The variability in column density suggests that the variations are caused by clouds moving in and out of the line of sight. Such changes in column density have also been observed in other sources, such as NGC~4151 \citep{Schurch2002,deRosa2007}, NGC~7582 \citep{Bianchi2009}, NGC~1365 \citep{Risaliti2005, Risaliti2009}. NGC~3822 can now be added to the list of sources that exhibit $N_{\rm H}$ variability. However, we need better quality observations with improved SNR to draw more promising conclusions.

\subsection{Optical spectral evolution of NGC~3822}
\label{sec:CLAGN3822}
Here, we present the evolution of the observed emission lines in the spectra of NGC~3822 from 2018 to 2025 based on optical spectroscopic observations from VLT and HCT. The first spectrum of NGC~3822 was taken in 1993. At that time, no broad emission lines (BELs) were present in the spectrum. Therefore, \cite{Moran1994} classified NGC~3822 as a Seyfert~2 AGN. In 1994, the source moved to a type~1.9 state with the appearance of BEL \halpha~\citep{Moran1996}. During 2018-2025, the line profiles are observed to change. Figure~\ref{fig:Hlpha_multiplot} illustrates the detailed evolution of both the \halpha~and \hbeta~spectral regions.  

In 1996, \citet{Moran1996} detected narrow \hbeta~ emission lines in the spectra of NGC~3822 with FWHM of a few hundred \kms and classified the source as type~1.9. In the 2018 XShooter spectrum (present work), we detected a narrow \hbeta~emission line with FWHM of $ 444\pm107$\kms. Based on our detection, NGC~3822 is also found to exhibit type~1.9 characteristics in 2018. After a four-year gap, in March 2022, \cite{Hinkle2022} classified the AGN as Seyfert~1 type, based on the appearance of BELs in the spectrum. The strong and broad Balmer lines that appeared in March 2022 remained there for subsequent Xshooter observations on 17 June and 11 July 2022 (Figure~\ref{fig:Hlpha_multiplot}). Both the \hbeta~and \halpha~regions required broad components with FWHM values exceeding 3500 \kms. Therefore, during this five-month observation period spanning from March 2022 to July 2022, NGC~3822 consistently exhibited the spectral characteristics of a Seyfert~1 type. Next, we monitored NGC~3822 with HCT from March 2024 to March 2025. During this one-year period, the broad \hbeta~and \halpha~emission lines were no longer present in the source spectra. Comparison of observations before and after 2022 shows that BELs appeared only in 2022 and disappeared again by 2024-2025. From the evolution of the line profiles, it is evident that NGC~3822 exhibits a changing-look behavior over the observation period from 2018 to 2025.

Considering the broader timeline from 1995 to 2025, we conclude that the source has undergone repeated changing-look transitions over this 30-year period. In 1993, the source was classified as type~2, and one year later, it transitioned to type~1.9. Although there were no spectroscopic observations between 1996 and 2018, the observation in 2018 revealed that the source remained in the type~1.9 state. After a four-year gap, in 2022, the source transitioned to spectral type~1. However, recent one-year optical monitoring with HCT, from 2024 to 2025, indicates that NGC~3822 gradually returned to the spectral type of 1.9/2. The spectral types observed at different epochs are illustrated in Figure~\ref{fig:Spectral_type}. Some of the CL-AGNs that show the repeated appearance and disappearance of broad emission lines are NGC~4151 \citep{Puccetti2007}, Mrk~1018 \citep{McElroy2016}, NGC~1365 \citep{Braito2014,Temple2023}, SDSS~J151652.48+295413.4 \citep{Lyu2025}. It is still uncertain whether every CL-AGN experiences repeated changing-look events, and the physical mechanisms that cause these events are unclear as well.

\subsubsection{Origin of the Changing-look behavior in NGC~3822}
In recent years, several dozen CL-AGNs have been discovered at different wavelengths. The physical mechanism for triggering the CL phenomenon in AGNs is still unclear. Some possible explanations of CL events are TDEs, a variation in the mass accretion rate, and changes in obscuration. In this work, we observed the AGN NGC~3822 across multiple epochs in both the X-ray and optical wavebands, providing us with information on how this source evolves over time. From the X-ray spectral study of NGC~3822, we identified variability in obscuration caused by clumpy clouds moving in and out of the line of sight. Long-term optical observations reveal that the source undergoes changes in the optical spectral state, transitioning from type~2$\rightarrow$1 and again type~1$\rightarrow$1.9. These transitions are caused by the appearance and disappearance of broad Blamer line profiles. Based on X-ray and optical observations, we conclude that NGC~3822 exhibits changing-look behavior in both X-ray and optical regimes. In the recent study of CL-AGNs, \cite{jana2025} investigated a possible correlation between obscuration and optical spectral state for a sample of AGNs. Their results indicate that, for the majority of the sample, the changes in the spectral state are not linked to variations in obscuration. However, a significant correlation was found between the optical spectral state and the Eddington ratio. Most of the type~1 CL-AGNs are associated with higher accretion rates, while AGNs transitioned to spectral type~2 correspond to relatively lower accretion rates \citep{jana2025}. For NGC~3822, we have three nearly coordinated X-ray and optical spectroscopic observations from the years 1993, 2022, and 2024. These observations provide an opportunity to examine changes in the spectral type along with the accretion rate. The spectral types at different observational epochs along with the Eddington ratio ($\rm log\lambda_{Edd}$) and $N_{\rm H}$ are shown in Figure~\ref{fig:Spectral_type}. The $\rm log\lambda_{Edd}$ for the year 1993 is calculated from the 0.1-2.4 keV luminosity reported by~\cite{Moran1994} using the bolometric correction factor $17.4\pm0.3$ taken from~\cite{Gupta2024}. The figure shows that when the source was in type~2 state (in 1993), the accretion rate was $\sim1.5\times10^{-3}$. However, when the source was in type~1 state (in 2022), the accretion rate was relatively high ($\sim3.80\times10^{-3}$). During 2024 to 2025 optical observations, the source transitioned to a type~2, accompanied by a significantly lower accretion rate of $\sim0.6\times10^{-4}$. The sudden changes in the accretion rate in 2022 that caused the CS event could be triggered by TDE. We observed a decline in UV flux consistent with the expected $t^{-5/12}$ trend for TDE (Figure~\ref{fig:TDE}). Based on observations spanning 1993 to 2024, we conclude that the changing-look phenomenon in NGC~3822 is primarily associated with changes in the accretion rate, rather than being caused by variable obscuration. However, to reach more robust conclusions, additional coordinated observations across X-ray and optical wavebands are required. 
\section{Summary and Conclusions}
\label{sec:conclusions}
For the first time, we conducted a multiwavelength study of the CL-AGN NGC~3822. We acquired data from the \swift, \xmm, \nustar, HCT, and VLT observatories during 2008--2025. Our findings are summarized below.

\begin{enumerate}

\item[1.] From the $\sim$16 years (2008--2024) of X-ray observations of NGC~3822, it is noted that the X-ray continuum luminosity (2--10 keV) of the source varies between $1.30\times10^{42}$ to $1.40\times10^{43}$\ergsec  and the corresponding Eddington ratio changes from$\sim0.8\times10^{-3}$ to $\sim9\times10^{-3}$. Based on the calculated Eddington ratio, it is evident that the source remained in a sub-Eddington regime during this observation period.

 \item[2.] The optical, ultraviolet, and X-ray light curves show a pattern of variability. The variability amplitude $F_{\rm var}$ decreases with increasing wavelength. Based on long-term monitoring (2008-2024), we report $\rm F_{\rm var}$ in the X-ray band as $\sim$63$\%$, which gradually decreases in the UV bands: $\sim$40$\%$ in W2, $\sim$39$\%$ in M2 and $\sim$33$\%$ in W1, followed by a further drop to $\sim29\%$ in the optical U band. In the optical bands, V and B, the variability is significantly lower with $\rm F_{var}$ $\sim$5-3$\%$. A similar trend of decreasing variability with increasing wavelength is also evident in the short-term (2022) monitoring campaign.

 \item[3.] Long-term X-ray monitoring reveals variability in the obscuring absorber of NGC~3822. The absorber was clearly detected during the 2016 and 2022 observations, while it disappeared before and after these epochs. The presence and absence of the absorber are caused by clouds moving in and out of the line of sight. 

 \item[4.] From the broadband X-ray spectral analysis, we found that the spectrum is well described by the Comptonization model CompTT. The estimated electron temperature of the corona lies between $50^{+21}_{-19}$ and $88^{+51}_{-32}$ keV. The optical depth of the corona is estimated to be $0.97^{+0.57}_{-0.34}$ and $0.77^{+0.46}_{-0.31}$.

  \item[5.] We carried out a detailed analysis of the 2022 flare, which follows a fallback trend consistent with a TDE. The UV flux decays as $t^{-0.41\pm0.03}$, which closely matches the theoretical $t^{-5/12}$ prediction expected from the thermal emission of a cooling accretion disc. This suggests that the observed flare in 2022 was likely triggered by a TDE event in the nucleus of NGC~3822.
  
  \item[6.] The change in accretion rate observed during the 2022 observation could be driven by the TDE, indicating a connection between the enhancement of the accretion rate and the TDE.

  \item[7.] NGC~3822 exhibits CL behavior in the optical band, characterized by the appearance and disappearance of BEL during the observation period from 2018 to 2025. 

  \item[8.] The CL transitions are found to be driven by the change in the accretion rate. The BELs are found at $\lambda_{\rm Edd}>3.8\times10^{-3}$, while the BELs disappear below this Eddington ratio. 
 
  \end{enumerate}
\section*{Acknowledgements}
We sincerely thank the reviewer for providing insightful comments and constructive suggestions that helped us improve the manuscript. The research work at the Physical Research Laboratory, Ahmedabad, is funded by the Department of Space, Government of India. AJ acknowledges support from the FONDECYT postdoctoral fellowship (3230303). The authors thank Prof. Michele Cappellari (University of Oxford) and Sorya Lambert (IEA/UDP) for their invaluable discussion on pPXF. The data and the software used in this work were taken from the High Energy Astrophysics Science Archive Research Center (HEASARC), which is a service of the Astrophysics Science Division at NASA/GSFC and the High Energy Astrophysics Division of the Smithsonian Astrophysical Observatory. This work has made use of data obtained from the \nustar~mission, a project led by Caltech, funded by NASA and managed by NASA/JPL, and has utilized the NuSTARDAS software package, jointly developed by ASDC, Italy, and Caltech, USA. This work made use of data \swift~supplied by the UK Swift Science Data Centre at the University of Leicester. We acknowledge the use of public data from the Swift data archive and thank the Swift team for approving the Swift TOO request. This research has used observations obtained with \xmm, an ESA science mission with instruments and contributions directly funded by ESA Member States and NASA. Based on observations collected at the European Organization for Astronomical Research in the Southern Hemisphere under the ESO programme(s) 0101.A-0765(A) and 109.24E8.001. The authors thank the staff of IAO, Hanle, and CREST, Hosakote, who made these observations possible. The facilities at IAO and CREST are operated by the Indian Institute of Astrophysics, Bangalore.
\section*{Data Availability}
We used optical data from ESO and IAO observatories in this work. The optical data from IAO can be shared on request and ESO data are publicly available. We used publicly available archival data from Swift, XMM–Newton and NuSTAR observatories in this work. These data are available on the corresponding websites. The appropriate links are given in the text.


\vspace{5mm}
\facilities{HEASARC,~\nustar,~\xmm,~\swift, ESO and IAO observatories}
\software{HEASOFT (v6.30),\texttt{XSPEC} (v-12.12.1, \citet{1996ASPC..101...17A}), IRAF~\citep{Tody1986}}
\bibliography{NGC3822}{}
\bibliographystyle{aasjournal}
\vspace{5mm}

\appendix
\restartappendixnumbering
\section{Optical/UV and X-ray fluxes}
\begin{table}[h]
\scriptsize
\renewcommand{\arraystretch}{1.0} 
    \centering
	\caption{Optical/UV monochromatic continuum flux taken from~\swift/UVOT and~\xmm/OM. The fluxes are in the unit of $10^{-15}$\efluxA.} 
	\label{tab:UV/optical_flux}
    \hspace*{-3.5cm}
	\begin{tabular}{l c c c c c c c c c c c c r} 
	\hline
     \hline
Instrument &Date      &MJD   &V            &B            &U            &UVW1         &UVM2        &UVW2\\ 
\hline
\swift/UVOT&2008-07-23&54670&--           &--           &--           &--           &$1.91\pm0.06$&--\\
\swift/UVOT&2009-07-28&55040&--           &--           &$2.44\pm0.05$&--           &--           &--\\
\XMM/OM    &2010-06-01&55348&--           &--           &--           &--           &$2.59\pm0.05$&--\\
\swift/UVOT&2013-07-26&56499&--           &--           &--           &$4.86\pm0.08$&--           &--\\
\swift/UVOT&2014-11-20&56981&--           &--           &$0.31\pm0.02$&--           &--           &$0.72\pm0.04$\\
\swift/UVOT&2015-03-24&57105&--           &--           &--           &--           &--           &$2.16\pm0.05$\\
\swift/UVOT&2015-07-17&57220&--           &--           &$1.20\pm0.03$&--           &--           &--\\
\swift/UVOT&2016-01-12&57399&--           &--           &--           &$1.72\pm0.06$&--           &--\\
\swift/UVOT&2016-01-15&57402&$6.66\pm0.21$&$5.38\pm0.14$&$2.61\pm0.09$&$1.72\pm0.07$&$1.42\pm0.07$&$1.48\pm0.06$\\
\swift/UVOT&2021-01-10&59224&--          &--           &$2.32\pm0.06$&--           &--            &--\\
\swift/UVOT&2021-07-27&59422&--          &--           &--           &--           &$1.17\pm0.06$ &--\\
\swift/UVOT&2021-07-28&59423&--          &--           &--           &$1.60\pm0.04$&--            &--\\
\swift/UVOT&2021-11-21&59539&--          &--           &--           &$1.58\pm0.07$&--            &--\\
\swift/UVOT&2022-03-30&59668&$6.92\pm0.28$&$6.01\pm0.20$&$3.69\pm0.15$&$3.33\pm0.12$&$2.98\pm0.10$&$2.95\pm0.10$\\
\swift/UVOT&2022-04-02&59671&--           &$6.17\pm0.15$&$4.05\pm0.11$&$3.61\pm0.09$&--           &$4.02\pm0.71$\\
\swift/UVOT&2022-04-10&59679&$7.10\pm0.21$&$5.82\pm0.15$&$3.81\pm0.11$&$3.17\pm0.09$&$2.74\pm0.07$&$2.54\pm0.07$\\
\swift/UVOT&2022-04-16&59685&$6.32\pm0.21$&$6.12\pm0.16$&$4.15\pm0.12$&$1.84\pm0.07$&$3.41\pm0.08$&$3.46\pm 0.09$\\
\swift/UVOT&2022-05-28&59727&$6.69\pm 0.21$&$ 5.63\pm 0.15$&$ 3.30\pm 0.10$&$ 2.58\pm 0.08$&$ 2.01\pm 0.06$&$1.99\pm 0.06$\\
\swift/UVOT&2022-06-12&59742&$6.66\pm 0.22$&$ 5.45\pm 0.15$&$ 3.24\pm 0.11$&$ 2.51\pm 0.08$&$ 2.19\pm 0.07$&$ 2.18\pm 0.07$\\
\swift/UVOT&2022-06-17&59747&$7.10\pm 0.28$&$ 5.65\pm 0.19$&$ 3.35\pm 0.14$&$ 2.89\pm 0.11$&$ 2.13\pm 0.08$&$ 2.39\pm 0.09$\\
\swift/UVOT&2022-06-26&59756&$7.02\pm 0.28$&$ 5.63\pm 0.19$&$ 3.56\pm 0.15$&$ 2.67\pm 0.11$&$ 1.83\pm 0.07$&$ 1.95\pm 0.08$\\
\swift/UVOT&2022-07-07&59767&$6.72\pm 0.22$&$ 5.51\pm 0.15$&$ 3.30\pm 0.11$&$ 2.66\pm 0.08$&$ 2.48\pm 0.07$&$ 2.19\pm 0.07$\\
\swift/UVOT&2022-07-17&59777&$6.46\pm 0.20$&$ 5.11\pm 0.14$&$ 3.34\pm 0.11$&$ 2.75\pm 0.08$&$ 2.22\pm 0.06$&$ 2.31\pm 0.07$\\
\swift/UVOT&2022-11-02&59885&$6.38\pm 0.18$&$ 5.35\pm 0.12$&$ 3.24\pm 0.09$&$ 2.43\pm 0.07$&$ 1.73\pm 0.07$&$ 1.93\pm 0.06$\\
\swift/UVOT&2022-11-09&59892&$6.07\pm 0.27$&$ 5.28\pm 0.19$&$ 3.29\pm 0.14$&$ 2.48\pm 0.13$&$ 2.24\pm 0.12$&$ 1.99\pm 0.09$\\
\swift/UVOT&2022-11-13&59896&--         &$5.27\pm0.11$&$3.34\pm0.08$&$2.97\pm0.07$&--             &$2.52\pm0.06$\\
\swift/UVOT&2022-11-20&59903&$7.10\pm0.19$&$5.64\pm0.13$&$3.41\pm0.09$&$2.86\pm0.08$&$2.53\pm0.11$&$2.52\pm0.06$\\
\swift/UVOT&2022-11-22&59905&$6.56\pm0.24$&$5.24\pm0.16$&$3.40\pm0.13$&$2.47\pm0.11$&$2.15\pm0.12$&$2.28\pm0.09$\\
\swift/UVOT&2022-11-29&59912&$6.47\pm0.19$&$5.28\pm0.13$&$2.81\pm0.09$&$2.13\pm0.08$&$1.65\pm0.07$&$1.76\pm0.06$\\
\swift/UVOT&2023-01-21&59965&--         &--           &--           &--           &--           &$1.37\pm0.09$\\
\swift/UVOT&2023-06-12&60107&$6.21\pm0.20$&$4.75\pm0.13$&$2.22\pm0.08$&$1.56\pm0.06$&$0.89\pm0.04$&$1.11\pm0.05$\\
\swift/UVOT&2023-07-13&60138&--         &--           &--           &--           &$1.12\pm0.04$&--\\
\swift/UVOT&2023-07-17&60142&--         &--           &--           &--           &$1.09\pm0.06$&--\\
\swift/UVOT&2023-07-18&60143&--         &--           &--           &$1.68\pm0.11$&--           &--\\
\swift/UVOT&2023-07-24&60149&--         &--           &--           &--           &--           &$1.17\pm0.03$\\
\swift/UVOT&2023-11-11&60259&--         &--           &--           &$1.48\pm0.04$&--           &--\\
\swift/UVOT&2023-12-09&60287&--         &--           &--           &--           &$0.95\pm0.04$&--\\
\swift/UVOT&2023-12-10&60288&--         &--           &--           &--           &$1.01\pm0.05$&--\\
\swift/UVOT&2023-12-22&60300&--         &--           &--           &--           &$1.03\pm0.03$&--\\
\swift/UVOT&2024-04-14&60414&--         &--           &--           &$1.64\pm0.10$&$0.89\pm0.08$&$0.86\pm0.08$\\
\swift/UVOT&2024-05-02&60432&--         &--           &--           &$1.65\pm0.05$&$1.15\pm0.04$&$1.33\pm0.05$\\
\swift/UVOT&2024-05-13&60443&--         &--           &--           &--           &$0.88 \pm0.04$&$0.78\pm0.03$\\
\swift/UVOT&2024-05-20&60450&--         &--           &--           &$1.72\pm0.06$&$1.20\pm0.06$&$1.27\pm0.06$\\
\swift/UVOT&2024-06-02&60463&--         &--           &--           &$1.58\pm0.04$&$1.14\pm0.04$&$1.18\pm0.04$\\
\swift/UVOT&2024-06-16&60477&--         &--           &--           &$1.48\pm0.04$&$1.17\pm0.04$&$1.30\pm0.04$\\
\swift/UVOT&2024-06-27&60488&--         &--           &--           &$1.46\pm0.04$&$1.30\pm0.05$&$1.25\pm0.05$\\

\hline
     \end{tabular}
\end{table}
\begin{table}[h]
\scriptsize
\renewcommand{\arraystretch}{1.0} 
    \centering
	\caption{ Unobscured X-ray fluxes in different energy bands from X-ray observations spanning 2008 to 2024. The fluxes are calculated using a simple power-law model fitting and fit statistics used Cash statistics~\citep{Cash1979}. All the X-ray fluxes are in the unit of~$10^{-12}$~\flux.}
        \label{tab:X_ray_flux}
        \hspace*{-3.5cm}
	\begin{tabular}{l c c c c c c c c c c c c c c r} 
	\hline
     \hline
Instrument &Date      &MJD  &$\Gamma$&$\rm Stat/dof$&$\rm F_{0.3-2 keV}$           &$\rm F_{2-10 keV} $&$\rm F_{0.3-10 keV}$\\ 
\hline
\swift/XRT &2008-07-23 &54670  &$1.54\pm0.18$&23.83/22    &$2.27\pm0.32$&$4.22\pm1.10$&$6.48\pm1.07$\\
\XMM/Epic-pn&2010-06-01&55348  &$1.54\pm0.02$&405.44/375  &$2.34\pm0.03$&$4.42\pm0.15$&$6.75\pm0.18$\\
\swift/XRT  &2013-07-26&56499  &$1.53\pm0.15$&55.19/49    &$8.16\pm0.99$&$15.56\pm3.00$&$23.71\pm3.03$\\
\swift/XRT  &2015-03-24&57105  &$1.30\pm0.47$&7.86/7      &$2.17\pm0.56$&$6.27\pm3.17$&$8.44\pm3.13$\\
\swift/XRT  &2016-01-12&57399  &$1.71\pm0.28$&8.45/14     &$2.65\pm0.57$&$3.72\pm1.11$&$6.38\pm1.20$\\
\swift/XRT  &2016-01-15&57402  &$1.80\pm0.24$&24.20/19    &$4.41\pm0.39$&$5.30\pm1.38$&$9.71\pm1.47$\\
\swift/XRT  &2022-03-30&59668  &$1.63\pm0.32$&9.55/9      &$3.12\pm0.91$&$5.03\pm1.62$&$8.15 \pm1.68$\\
\swift/XRT  &2022-04-02&59671  &$1.97\pm0.32$&13.71/10    &$7.06\pm1.86$&$6.24\pm2.14$&$13.29\pm2.22$\\
\swift/XRT  &2022-04-10&59679  &$2.10\pm0.14$&40.04/48    &$8.80\pm1.16$&$6.16\pm1.10$&$14.97\pm1.48$\\
\swift/XRT  &2022-04-16&59685  &$1.78\pm0.18$&26.89/30    &$4.89\pm0.89$&$6.18\pm1.26$&$11.08\pm1.38$\\
\swift/XRT  &2022-05-28&59727  &$1.64\pm0.22$&27.08/25    &$5.97\pm1.21$&$9.54\pm2.16$&$15.52\pm2.13$\\
\swift/XRT  &2022-06-12&59742  &$1.74\pm0.16$&54.03/41    &$5.70\pm0.92$&$7.63\pm1.31$&$13.32\pm1.42$\\
\swift/XRT  &2022-06-17&59747  &$1.35\pm0.32$&11.22/10    &$2.90\pm0.86$&$7.52\pm2.24$&$10.42\pm2.26$\\
\swift/XRT  &2022-06-26&59756  &$1.78\pm0.41$&5.03/6      &$2.53\pm0.28$&$3.10\pm0.80$&$5.70 \pm1.41$\\
\swift/XRT  &2022-07-07&59767  &$1.80^{f}$   &14.84/11    &$4.59\pm0.13$&$5.24\pm1.36$&$11.97\pm1.33$\\
\swift/XRT  &2022-07-17&59777  &$1.80\pm0.14$&46.92/52    &$5.60\pm0.24$&$6.66\pm1.79$&$12.26\pm1.16$\\
\swift/XRT  &2022-11-02&59885  &$1.50\pm0.13$&54.51/62    &$4.94\pm0.50$&$10.04\pm1.59$&$15.00\pm1.24$\\
\swift/XRT  &2022-11-09&59892  &$1.53\pm0.13$&45.75/49    &$3.56\pm0.43$&$6.76\pm1.26$&$10.32\pm1.29$\\
\swift/XRT  &2022-11-13&59896  &$1.52\pm0.16$&24.23/39    &$5.57\pm0.64$&$10.81\pm2.21$&$16.37\pm2.24$\\
\swift/XRT  &2022-11-20&59903  &$1.56\pm0.14$&45.20/52    &$5.19\pm0.59$&$9.40\pm1.71$&$14.60\pm1.75$\\
\swift/XRT  &2022-11-22&59905  &$1.52\pm0.12$&93.33/71    &$3.67\pm0.37$&$7.08\pm1.08$&$10.75\pm1.10$\\
\swift/XRT  &2023-06-12&60107  &$1.30\pm0.48$&18.93/13    &$0.30\pm0.11$&$0.81\pm0.39$&$1.11\pm0.40$\\
\swift/XRT  &2023-07-13&60138  &$1.34\pm0.31$&12.84/18    &$0.84\pm0.21$&$2.20\pm0.76$&$3.05\pm0.78$\\
\swift/XRT  &2023-07-24&60149  &$1.40\pm0.54$&9.39/7      &$0.70\pm0.25$&$1.47\pm0.80$&$2.17\pm0.81$\\
\swift/XRT  &2023-12-22&60300  &$1.48\pm0.74$&1.51/4      &$0.33\pm0.17$&$0.69\pm0.46$&$1.03\pm0.48$&\\
\swift/XRT  &2024-04-14&60414  &$1.95\pm0.62$&2.26/4      &$0.84\pm0.24$&$0.77\pm0.50$&$1.62\pm0.55$&\\
\swift/XRT  &2024-05-02&60432  &$1.30\pm0.33$&12.29/9     &$0.68\pm0.19$&$1.90\pm0.70$&$2.57\pm0.70$\\
\swift/XRT  &2024-05-13&60443  &$1.43\pm0.41$&7.86/8      &$1.00\pm0.16$&$2.27\pm0.96$&$3.27\pm0.93$\\
\swift/XRT  &2024-05-20&60450  &$1.88\pm0.54$&5.06/4      &$0.89\pm0.25$&$0.94\pm0.32$&$1.82\pm0.58$\\
\swift/XRT  &2024-06-02&60463  &$1.73\pm0.60$&2.34/3      &$0.38\pm0.18$&$0.53\pm0.32$&$0.90\pm0.36$&\\
\swift/XRT  &2024-06-16&60477  &$1.58\pm0.36$&7.38/8      &$0.68\pm0.10$&$1.17\pm0.47$&$1.85\pm0.47$\\
\swift/XRT  &2024-06-27&60488  &$1.48\pm0.20$&25.56/26    &$1.54\pm0.23$&$3.23\pm1.01$&$4.77\pm1.00$\\
\swift/XRT  &2024-07-13&60504  &$1.48^{f}   $&6.71/8      &$0.66\pm0.14$&$1.38\pm0.29$&$2.04\pm0.35$\\
\hline
     \end{tabular}
\end{table}
\label{lastpage}
\end{document}